\documentclass[twocolumn,prx,aps,superscriptaddress,longbibliography]{revtex4-2}
\usepackage{amssymb,amsmath,amsfonts}

\usepackage{url}
\usepackage[latin9]{inputenc}
\usepackage{multirow}

\usepackage{amstext}
\usepackage{braket}
\usepackage{enumitem}
\usepackage{graphicx,bm,palatino}
\usepackage[colorlinks=true,linkcolor=blue,urlcolor=blue,citecolor=blue,pdfusetitle]{hyperref}

\usepackage{bm}

\usepackage[sc]{mathpazo} 

\usepackage{hyperref,cleveref}

\usepackage[dvipsnames]{xcolor}
\usepackage[caption=false]{subfig}

\usepackage{times}
\usepackage{bbm}
\makeatother

\begin{document}
\title{Machine Learning-Enhanced Characterisation of Structured Spectral Densities: Leveraging the Reaction Coordinate Mapping}
\date{\today}

\author{J. Barr}
\email{jbarr24@qub.ac.uk}
\affiliation{Centre for Quantum Materials and Technologies, School of Mathematics and Physics, Queen's University Belfast, BT7 1NN, United Kingdom}
\author{A. Ferraro}
\affiliation{Dipartimento di Fisica ``Aldo Pontremoli'', Universit\`{a} degli Studi di Milano, I-20133 Milano, Italy}
\author{M. Paternostro}
\affiliation{Quantum Theory Group, Dipartimento di Fisica e Chimica Emilio Segr\`e, Universit\`a degli Studi di Palermo, via Archirafi 36, I-90123 Palermo, Italy}
\affiliation{Centre for Quantum Materials and Technologies, School of Mathematics and Physics, Queen's University Belfast, BT7 1NN, United Kingdom}
\author{G. Zicari}
\email{G.Zicari@qub.ac.uk}
\affiliation{Centre for Quantum Materials and Technologies, School of Mathematics and Physics, Queen's University Belfast, BT7 1NN, United Kingdom}

\begin{abstract}
Spectral densities encode essential information about system-environment interactions in open-quantum systems, playing a pivotal role in shaping the system's dynamics. In this work, we leverage machine learning techniques to reconstruct key environmental features, going beyond the weak-coupling regime by simulating the system's dynamics using the reaction coordinate mapping. For a dissipative spin-boson model with a structured spectral density expressed as a sum of Lorentzian peaks, we demonstrate that the time evolution of a system observable can be used by a neural network to classify the spectral density as comprising one, two, or three Lorentzian peaks and accurately predict their central frequency. 
\end{abstract}

\maketitle

\section{Introduction}
\label{sec:intro}

{The translation of theoretical protocols for quantum information processing into actual quantum technologies require
very precise control over non-classical states and processes. However, quantum states are typically very fragile against environmental noise, the latter coming either in the form of decoherence -- which destroys quantum superposition -- or dissipation. The interaction of a quantum system with its environment is not necessarily detrimental, as it could be  exploited to enhance the performance of specific tasks. For instance, local dephasing can enhance excitation transport across the sites of a quantum network~\cite{Plenio:2008,Mohseni:2008,Rebentrost:2009,Caruso:2009,Sgroi:2024}, while  suitably engineered Markovian open-system dynamics can be leveraged to the scopes of quantum state engineering~\cite{Kraus:2008}. In fact, by resorting to controlled dissipation,
quantum reservoir engineering  can be successfully exploited to manipulate quantum information~\cite{Poyatos:1996,Verstraete:2009,Harrington:2022}, thus 
driving quantum systems towards specific target states or non-trivial phases of matters that exhibits interesting quantum features.}

{For such a class of tasks, knowledge of the system-environment interaction is key. In the {\it canonical} phrasing of 
open quantum dynamics, 
a typically small number of degrees of freedom of a quantum system are coupled to 
infinitely many (controllable or  uncontrollable) degrees of freedom with the environment~\cite{Breuer-Petruccione1}. Here, a crucial issue is where to place the boundary that sets the system and the environment apart. 
This choice is mirrored in the functional form of the so-called Spectral Density (SD), a function that encodes full information about the coupling between the system and the environment~\cite{deVEga:2017}. The SD allows us to determine the multi-time correlation functions of the environment, which, in turn, allows us to predict the evolution of open quantum systems without a complete microscopic description of their environment. In general, identifying the SD from first principles is a challenging task. In practice, the SD is inferred phenomenologically through empirical data or, in some special cases, approximated using \emph{ad hoc} assumptions. This often results in significant inaccuracies when predicting the system's dynamics~\cite{PhysRevLett.118.100401}.
}

To address this challenge, we tackle the problem of characterising environmental effects on an open quantum system by leveraging recent advances in the field of Machine Learning (ML). Such progress has enabled the introduction of data-driven approaches into the manipulation of quantum systems and processes that have proven to be highly effective in applications
such as quantum tomography~\cite{Palmieri2020, Torlai2018, Banchi_2018}, quantum channel discrimination~\cite{PhysRevA.106.032409},  quantum control~\cite{GIANNELLI2022128054, Niu2019,sgroi2021reinforcement,Brown:2021,khalid2023sample,Sgroi:2025}, as well as the simulation or characterisation of open quantum dynamics~\cite{doi:10.1126/science.aag23021, PhysRevLett.128.090501, BANDYOPADHYAY2018272,  PhysRevLett.122.160401, wei2024finding, nelson2022data,goswami2021experimental,mahlow2024predicting}.

{Usually, the partition between system and environment leads to SDs that give rise to a rapid decay of the bath correlation functions. This scenario is often compatible with the assumptions underlying the so-called Born-Markov approximation, which is a workhorse in the description of the dynamics of a quantum system weakly coupled to an external reservoir. In this regime, the environment can be treated as stationary, leading to Markovian evolutions characterised by a continuous and irreversible flow of information from the system to the environment~\cite{Gardiner:2009,CHRUSCINSKI20221}. However, certain choices of SD  invalidate such an approximation, leading to a backflow of information from the environment to the system~\cite{Rivas:20141,breuer2016colloquium}.}

To address these limitations, we turn to the reaction coordinate (RC) mapping~\cite{10.1063/1.449017, 10.1063/1.1385562, Nazir2018, 10.1063/1.3532408, Strasberg_2016, 10.1063/1.5114690, Puebla:2019}. This technique involves performing an exact transformation on the environment, which incorporates the environmental degrees of freedom that are responsible for strong coupling or non-Markovian effects into an effective enlarged system. The latter consists of the original system coupled to a collective coordinate of the environment, known as the RC, which is in turn coupled to a simpler, memoryless residual bath. The dynamics of such enlarged system is thus well described by a master equation in the Born-Markov approximation.

As a case study of enormous theoretical and practical interest, here we consider a dissipative spin-boson model, which describes a two-level system interacting with a bosonic thermal bath. The system's behaviour is strongly influenced by the features of the SD governing the system-environment interaction. Non-Markovian effects arise not only from strong system-environment coupling, but also when the SD is of a structured form, with some environmental modes being more strongly coupled to the system than others~\cite{breuer2016colloquium,Gribben:2020, gribben2022using, mouloudakis2022coalescence, ritschel2011efficient}. To capture such complex dynamics, we consider structured SDs modelled as a sum of Lorentzian peaks, which allows us to explore a wide range of coupling regimes, including scenarios with significant memory effects and strong system-environment interactions.
{The latter are very common in the study of light-harvesting complexes, such those customarily addressed  in the field of quantum biology~\cite{Lambert:2013,Huelga_Plenio_quantum_bio:2013}, where highly structured SDs often appear. For instance, in the celebrated Fenna-Matthews-Olson (FMO) photosynthetic complex the estimated phonon SD is given -- in the high frequency part of the spectrum -- by several peaks, modelled as a sum of narrow Lorentzian functions~\cite{meier1999non,kreisbeck2012long,Kreisbeck:2014,Caycedo-Soler:2022,Lorenzoni:2024}. Moreover, it has been recently shown that it is possible coarse-grain the environmental features through a systematic technique that allows to replace a highly structured SD with a simpler one where the number of peaks is dramatically reduced~\cite{Lorenzoni:2024}. Thanks to this, one is able to accurately describe the system dynamics within a finite time window, allowing to reproduce absorption spectra.}
{We consider cases where the system-bath interaction is modelled through a SD with a few Lorentzian peaks.}
We first demonstrate that an artificial neural network (NN) can be used to \emph{classify} the SD governing the dynamics of the system, based on the time evolution of a system observable. Specifically, we show that the NN can distinguish between dynamics characterised by one, two or three peaks. Previous research has explored the classification of various \emph{aspects} of noise in quantum systems. For instance, ML techniques have been used to discern between various Markovian and non-Markovian noise sources~\cite{martina2023machine,mukherjee2024noise}, or to classify the noise fingerprints of IBM quantum computers~\cite{martina2022learning}. Additionally, in Ref.~\cite{palmieri:2021}, tomographic data at just two instants of time were used to classify the environmental parameters, including the Ohmicity parameter, in single-qubit dephasing channels. Moreover, in a previous study~\cite{barr2024spectral}, we used the time evolution of an observable and NNs to classify the Ohmicity parameter in spin-boson models that were either simple enough to be solved exactly, or restricted to the weak coupling regime. 

In this study, we explore the case of more structured and realistic SDs instead. After categorising the dynamics based on the number of Lorentzian peaks present in the SD, we proceed with a regression task to estimate the frequency associated with these peaks. Prior studies have explored the use of regression analysis for noise characterisation in open quantum systems. For example, some works have investigated the noise in qubit systems using two-pulse echo decay curves~\cite{PRXQuantum.2.010316}, and random pulse sequences that are applied to the qubit~\cite{Youssry2020}. Other research has focused on constructing the power spectral density for ensembles of carbon impurities around a nitrogen vacancy centre in diamond~\cite{martina2023deep}, characterising noise parameters in neutral atoms devices~\cite{canonici2024machine}, and estimating the noise exponent and amplitude of $1/f$ noise in spin-qubit devices~\cite{zhang2019spin}. Additionally, Ref.~\cite{papivc2022neural2} presents an approach in which both classification and regression are used to infer the environment affecting superconducting qubits, while Ref.~\cite{garau2019machine2} focuses on inferring the Ohmicity class and performing regression on the Ohmicity parameter and damping strength, using a probing scheme that leverages the special features of quantum synchronization.

To gain further insights into the factors contributing to high classification accuracy and those that may introduce challenges, we analyse correlations between the predicted classification probabilities and specific features of the trajectories, such as the values of the parameters in the SD, peak spacing, and the distances of the peaks from the bare oscillator frequency. Additionally, we introduce a forced separation between peaks and demonstrate that classification accuracy improves as the peak separation increases. We also discuss how peak separation can enhance prediction accuracy in the regression tasks. 

The remainder of this paper is structured as follows: in~\Cref{sec:methods} we provide an introduction to the general setting under consideration, the RC mapping, and the ML approach employed. Specifically, we examine a two-level system interacting with a bosonic environment and provide some background on NNs. In \Cref{sec:results} we discuss NN architecture for the classification and regression tasks, along with a detailed analysis of the training and testing for each model. We present our concluding remarks and future outlook in \Cref{sec:conclusions}, while we present a number of technical results in a set of Appendices.

\section{Context and Methodology}
\label{sec:methods}

\begin{figure}
\centering\includegraphics[width=\columnwidth]{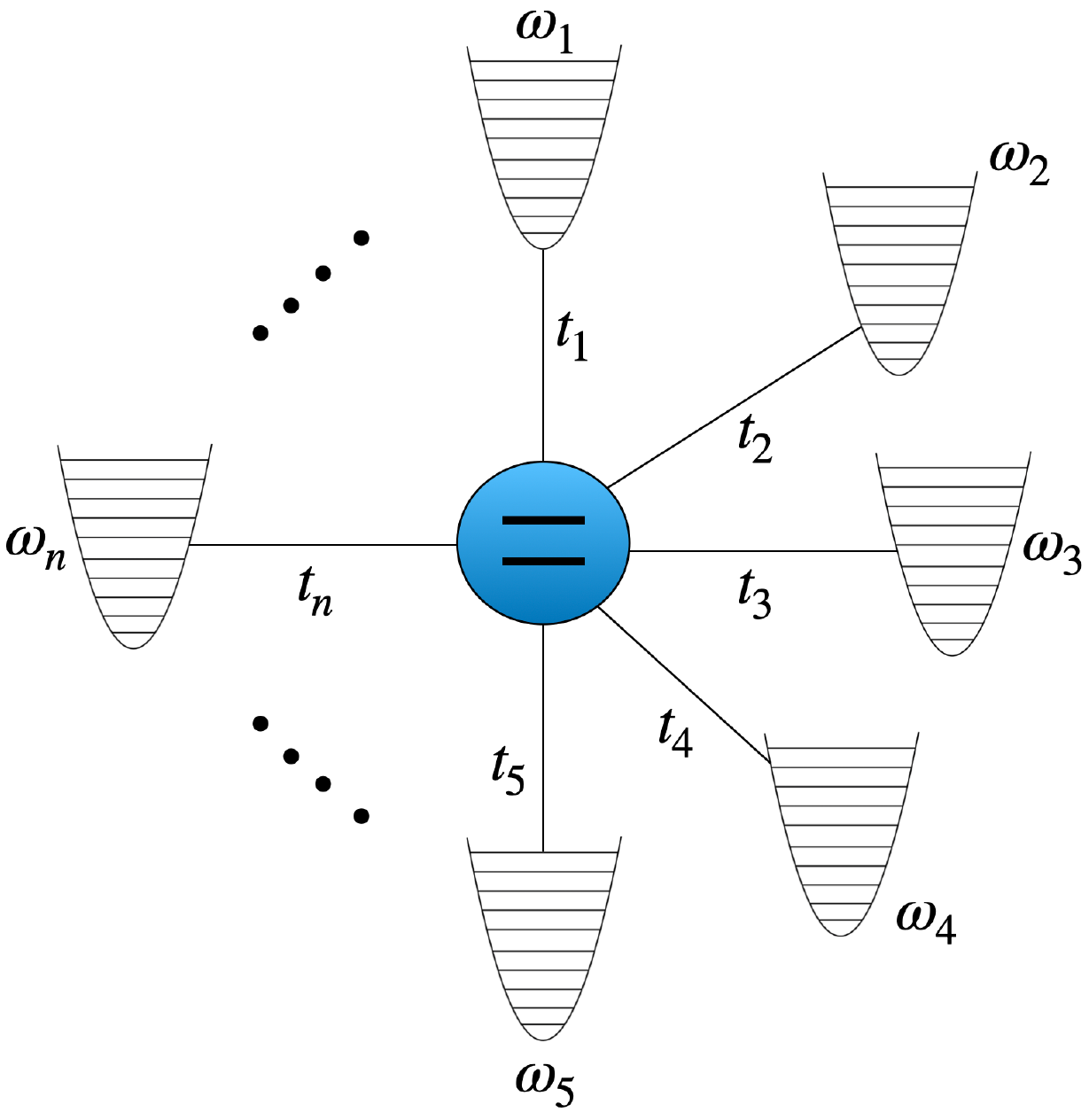}
	\caption{A two-level system interacting with a bosonic environment composed of an infinite set of harmonic oscillators, labelled by the integer $n$. The oscillators are characterised by their respective frequencies $\omega_n$ and couple to the two-level system with coupling strengths $t_n$.}\label{fig:SBmodel}
\end{figure}

Let us consider a spin-boson model, which describes a two-level system interacting with an environment composed of infinitely many bosonic harmonic oscillators, as shown in \Cref{fig:SBmodel}. The Hamiltonian of the spin-boson model can be written as 
\begin{equation}\label{eq:Hamiltonian}
    H = H_S + H_E + H_I,
\end{equation}
where $H_I$ is the interaction Hamiltonian which describes the coupling between the two-level system and the bosonic environment and reads as (we use units such that $\hbar=1$ throughout the manuscript)
\begin{equation}\label{eq:H_I}
    H_I = X \otimes B \, ,
\end{equation}
where $X$ is a generic system-only operator, while $B$ is an operator of the bath. We take the latter as
\begin{equation}\label{eq:bath_operator}
    B = \sum_k \left( t_k a_k^{\dagger} + t_k^* a_k \right) \, .
\end{equation}
Here, $k$ labels the reservoir modes of frequency $\omega_k$, and $a_k^{\dagger}$ and $a_k$ are the bosonic creation and annihilation operators of mode $k$, respectively. The coefficient $t_k$ determines the interaction strength between the two-level system and the $k$-th mode. The free Hamiltonians of the system and the environment, $H_S$ and $H_E$, are given by 
\begin{align}
   H_S = \frac{\omega_0}{2} \sigma_z, 
   \quad H_E = \sum_k \omega_k a_k^\dagger a_k 
\end{align}
with $\sigma_z$ being the $z$-Pauli operator and $\omega_0$ being the level spacing of the two-level system. The coupling coefficients enter into the formal definition of the SD, $J (\omega) = \sum_k |t_k|^2 \delta (\omega - \omega_k)$, which essentially parameterises the coupling coefficients and encodes all the information about the system-environment coupling. We assume that the distribution of modes forms a continuum so to ensure that the system dynamics do not display recurrences~\cite{Breuer-Petruccione1,Rivas:2010, Pucci:2013}. In this limit, the SD appears in the expression for the correlation function of a bosonic bath, defined as $C (\tau) \equiv \langle \tilde{B}(\tau) \tilde{B}(0) \rangle_E$, where $\tilde{B}(\tau)$ is the bath operator in the interaction picture with respect to the free Hamiltonian $H_E$.  In Appendix \ref{app:A}, we calculate $C(\tau)$ when the environment is modelled as a large bosonic thermal reservoir, such that $\rho_E = e^{- \beta H_E}/ \mathcal{Z}_E$, where $\mathcal{Z}_E \equiv \operatorname{tr}_E (e^{- \beta H_E})$ is the reservoir partition function. It can be expressed as
\begin{equation}\label{eq:bosoniccorrfunc}
    C (\tau) = \int_0^{\infty} \textrm{d} \omega J (\omega) \left[   n_B (\omega)e^{i \omega \tau} + (1 + n_B (\omega))e^{-i \omega \tau} \right] \, ,
\end{equation}
where $n_B (\omega) = \left( e^{\beta \omega} - 1 \right)^{-1}$ is the Bose-Einstein distribution and $\beta = 1/T$, with $T$ being the temperature of the environment (in the chosen sets of units, we have set the Boltzmann constant $k_B=1$). The effective dynamics of the two-level system, governed by a master equation, fundamentally depend on the correlation function $C (\tau)$, which represents the fingerprint of the environment. Moreover, the function $C (\tau)$ is ultimately determined by the shape of the SD, which contains all of the information about the environment needed to solve the dynamics of the system, and, subsequently determine the time evolution of its observables. The expectation value of a generic observable $O$ of the system at time $t$ is indeed given by
\begin{align}
\label{eq:observable}
\langle O (t) \rangle \equiv \operatorname{tr}_{SE} \left ( O e^{- i H t} \, \rho_{SE} (0) \, e^{i H t}\right )
\end{align}
with $H$ as in \Cref{eq:Hamiltonian}, while the global initial state is factorised as $\rho_{SE} (0) = \rho_S (0) \otimes \rho_E$, with $\rho_S (0)$ and $\rho_E$ being the initial system and environmental states, respectively.  Under these conditions, it can be shown that the only environmental quantity entering in the expression for $\langle O (t) \rangle$ is the SD $J(\omega)$. 

Qualitatively, non-Markovian memory effects typically emerge at strong system-bath coupling, which can result from a SD exhibiting a peaked behaviour, such that specific environmental modes are more strongly coupled to the system than others, in which case we say that the SD is structured \cite{breuer2016colloquium, Gribben:2020, gribben2022using, mouloudakis2022coalescence, ritschel2011efficient}. A prime example of a structured SD is a Lorentzian-shaped SD such as 
\begin{equation}
    J^L_i (\omega) = \frac{\lambda_i^2 \gamma_i \omega}{\left(\omega^2 - \nu_i^2 \right)^2 + \gamma_i^2 \omega^2} \, ,
\end{equation}
which is peaked around a central frequency $\nu_i$ with a width $\gamma_i$ and entails a system-bath coupling strength $\lambda_i$. A Lorentzian-shaped spectral density can serve as a basis for representing a wide range of other spectral densities \cite{lambert2019modelling}. Ref.~\cite{meier1999non} introduced a scheme for fitting spectral densities to a sum of Lorentzian functions, which has successfully been applied to various problems, from fitting experimentally determined spectral densities for light-harvesting systems~\cite{kreisbeck2012long} to calculating linear absorption spectra~\cite{schroder2006calculation} and characterizing the time dependence of fluorescence anisotropy~\cite{kleinekathofer2003memory}. Moreover, this approach has been employed in simulating the dynamics of driven Brownian oscillators \cite{yan2005quantum}, and damped harmonic oscillators \cite{kleinekathofer2004non}, as well as in studying charge transfer dynamics at oligothiophene-fullerene heterojunctions \cite{10.1063/1.4861853}. To facilitate the characterisation of a wide variety of SDs, we consider those that can be represented as a sum of $N$ Lorentzian functions, and can thus be written as
\begin{equation}
    J (\omega) = \sum_{i=1}^N J_{i}^L (\omega) \, .
\end{equation}

\subsection{The Pure Dephasing and Amplitude Damping Models}
\label{sec:PDandAD}

When $X = \sigma_z$ in \Cref{eq:H_I}, the interaction Hamiltonian commutes with the system Hamiltonian. Consequently, the populations of the reduced density matrix are left invariant by the dynamics. In this particular case, it is possible to access the full unitary evolution, and exactly trace out the environmental degrees of freedom, leading to an analytical solution for the reduced dynamics~\cite{Breuer-Petruccione1, Palma:96, Guarnieri:2014}. In \Cref{app:B}, we derive a solution for the dynamics under the standard assumption that the system and the environment are initially uncorrelated, with the environment described by a thermal Gibbs state. Working in the Schr{\"o}dinger picture, the time-evolved reduced density matrix at time $t$ can be expressed in the $\sigma_z$ basis $\{ \ket{0}, \ket{1} \}$ as
\begin{equation} \label{eq:Exactlysolvablerhot}
    \rho_S (t) = \begin{pmatrix}
    \rho_{00} (0) & \rho_{01} (0)  e^{- \Gamma (t) - i \omega_0 t} \\
    \rho_{01}^{*} (0) e^{- \Gamma (t) + i \omega_0 t} & 1-\rho_{00} (0)
    \end{pmatrix},
\end{equation}
where the decoherence function is given by 
\begin{equation}\label{eq:decoherencefunction}
    \Gamma (t) = 4 \int_0^{\infty} \textrm{d} \omega J (\omega) \coth \left( \frac{\beta \omega}{2} \right) \frac{1 - \cos (\omega t)}{\omega^2} \, .
\end{equation}
As seen in \Cref{eq:Exactlysolvablerhot}, the interaction with the environment leads to pure dephasing in the $\sigma_z$ basis without any dissipation. Notably, certain forms of the SD can result in negative values of $\Gamma(t)$, during which the system exhibits \emph{recoherence}. This phenomenon arises due to (non-Markovian) memory effects inherent in the dynamics~\cite{Addis:2014}. 

Alternatively, we can explore a scenario beyond pure dephasing by selecting $X = \sigma_x$ in the interaction Hamiltonian of \Cref{eq:H_I}. Unlike the case where $X = \sigma_z$, this choice results in a Hamiltonian that does not exhibit any explicit symmetry, preventing us from obtaining an exact solution for the dynamics. However, we can still effectively solve the dynamics, provided that we rely on further assumptions. For systems weakly coupled to their environment, a commonly used approach is to derive a master equation within the Born-Markov approximation~\cite{Breuer-Petruccione1}, which -- in the Schr{\"o}dinger picture -- reads as 
\begin{equation}
\begin{aligned}\label{eq:BM}
    \frac{\partial \rho_S (t)}{\partial t} &=  - i \left[ H_S, \rho_S (t) \right]\\
    &- \int_0^{\infty} \textrm{d} \tau \left( C (\tau) \left[ X, X (- \tau) \rho_S (t) \right] +\text{h.c.}\right) 
\end{aligned}
\end{equation}
where $X (- \tau) = e^{-i H_S \tau} X e^{i H_S \tau}$, and $C(\tau)$ is the bath correlation function, as defined in \Cref{eq:bosoniccorrfunc} for a bosonic thermal reservoir and such that $C^*(\tau)=C(-\tau)$. \Cref{eq:BM} leads to Markovian dynamics, characterized by a continuous, irreversible flow of information from the system to the environment. 

The Born-Markov approximation is valid only under certain conditions, relying on the assumptions of weak system-environment coupling and a rapid decay of the bath correlation function, which makes the environment's memory effects negligible. However, certain SDs can render the Born-Markov approximation invalid. In such cases, the environment retains memory of its interaction with the system over extended periods, leading to non-Markovian dynamics where information flows back into the system from the environment. 
Consequently, the Born-Markov approximation fails to accurately describe the dynamics of the system under these conditions, necessitating the use of more sophisticated non-perturbative approaches that account for non-Markovian effects. One such approach is the RC mapping. 

\subsection{The Reaction Coordinate Mapping}

In the RC mapping~\cite{10.1063/1.449017, 10.1063/1.1385562, Nazir2018, 10.1063/1.3532408, Strasberg_2016, 10.1063/1.5114690,Puebla:2019}, we perform an exact transformation on the bath to incorporate the environmental degrees of freedom which are responsible for strong coupling or non-Markovian effects into an effective, enlarged system. The resulting enlarged system consists of the original system coupled to a collective coordinate of the environment, known as the RC, which is in turn coupled to a residual bath. This transformation aims to enable the derivation of a master equation for the enlarged system within the Born-Markov approximation. Originally, the coupling between the system and the environment was too strong to validate a Born-Markov approach. However, after the mapping, the strong coupling is incorporated into the enlarged system, leaving only weak coupling between the RC and the residual bath. This allows us to solve for the dynamics in a wider range of regimes. 

\begin{figure}
\centering\includegraphics[width=\columnwidth]{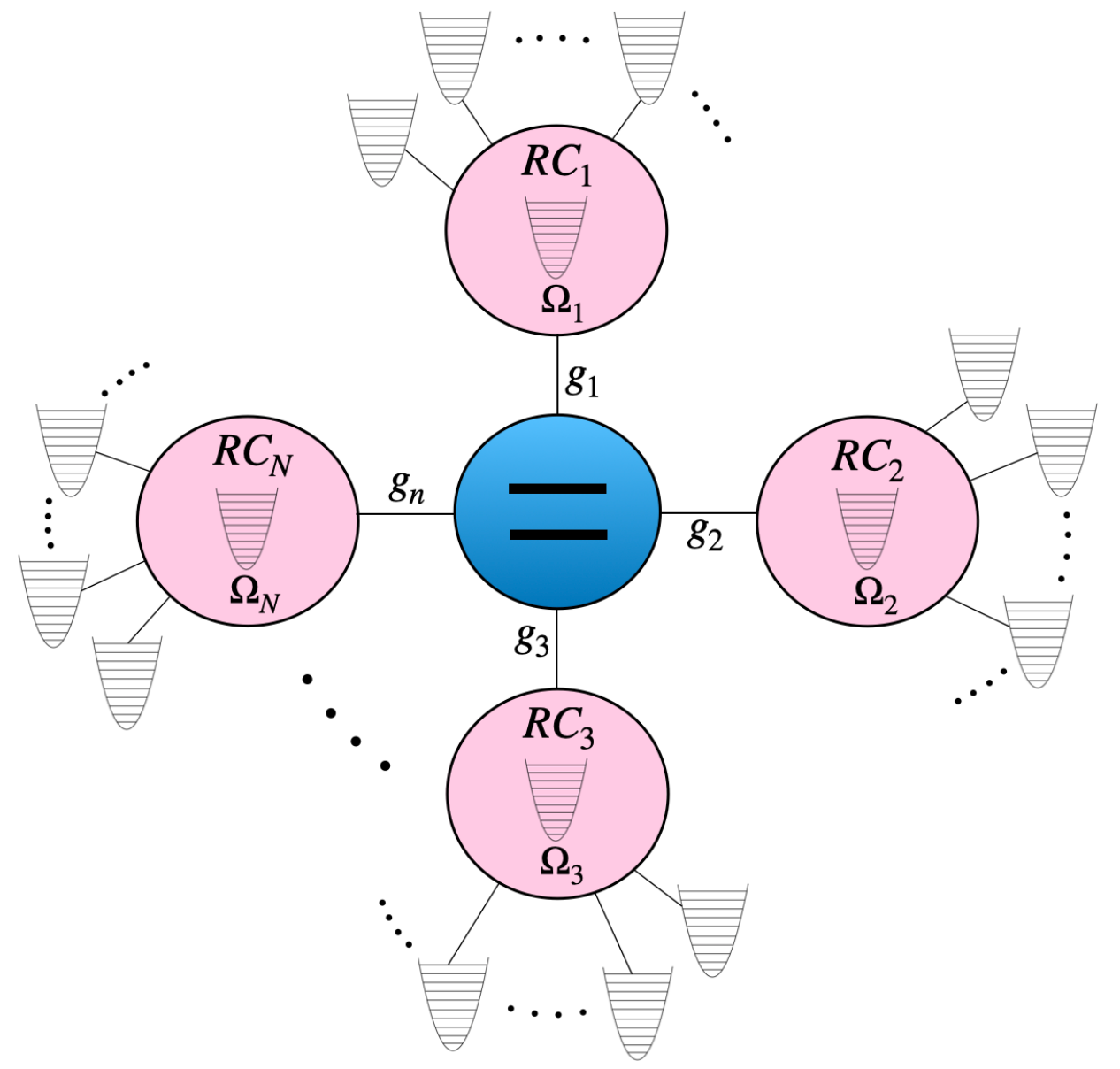}
\caption{Diagram of the spin-boson model after the RC mapping. The two-level system now interacts with a set of RCs, each labelled by an integer $n$. Every RC is characterized by a frequency $\Omega_n$ and interacts with the two-level system with a coupling strength $g_n$. Additionally, each RC is coupled to the residual environment.}\label{fig:modelaftermapping}
\end{figure}

Building on this technique, if the SD can be written as a sum of $N$ different contributions, one can transform the original environment and incorporate multiple collective coordinates into the augmented system, with each RC corresponding to a distinct contribution to the SD. After the mapping, each of these RCs interacts with the residual environment through a distinct SD, effectively resulting in each RC experiencing a different environment, as shown schematically in \Cref{fig:modelaftermapping}. Before the mapping, we have the spin-boson model depicted in \Cref{fig:SBmodel}, whose Hamiltonian is given in \Cref{eq:Hamiltonian}. The interaction between the system and the environment is characterized by the SD $J_0 (\omega) = \sum_k |t_k|^2 \delta (\omega - \omega_k)$. Post mapping, the Hamiltonian is written as
\begin{equation}\label{eq:Hamiltonianafter}
    H = H_{S'} + H_{E'} + \sum_{i \leq N} H_{{RCE}_i} \, ,
\end{equation}
where $H_{S'}$ is the free Hamiltonian of the effective enlarged system, $H_{E'}$ is the free Hamiltonian of the residual environment and $H_{{RCE}_i}$ describes the interaction between the $i$-th RC and the residual environment. Each term reads as 
\begin{equation}
\begin{aligned}
    H_{S'} &=  H_S + \sum_{i \leq N} \Omega_i b_i^{\dagger} b_i + \sum_{i\leq N} g_i X \left( b_i + b_i^{\dagger} \right) \\
    & + \sum_{i \leq N} \left[g_i^2\frac{ X^2}{\Omega_i} + \sum_{ k > N} \frac{h_{ik}^2}{\Omega_k} \left( b_{i k} + b_{i k}^{\dagger} \right)^2\right],\\
    H_{E'} &=  \sum_{i \leq N} \sum_{k > N} \Omega_k b_{ik}^{\dagger} b_{ik},\, \\
    H_{{RCE}_i}& = (b_i + b_i^{\dagger}) \sum_{k > N} h_{ik} (b_{ik} + b_{ik}^{\dagger}).
\end{aligned}
\end{equation}
Here, $\Omega_i$ and $\Omega_k$ represent the frequencies of the RCs and the $k$-th residual bath mode respectively, while $b_i$, $b_i^{\dagger}$, $b_{ik}$, and $b_{ik}^{\dagger}$ denote their corresponding annihilation and creation operators. The coupling strengths $g_i$ determine the interaction between the two-level system and the $i^\text{th}$ RC, and the coupling coefficients $h_{ik}$ determine the interaction strength between the $i$-th RC and the residual environment. {The mapping is realised through a Bogoliubov transformation, whereby bosonic creation and annihilation operators are linearly mapped to new operators 
\begin{equation}\label{eq:Bogoliubovtransformation}
    a_k  = \sum_{p \leq N} \left( u_{kp} b_p + v_{kp} b_p^{\dagger} \right) + \sum_{q > N} \left( u_{kq} b_q + v_{kq} b_q^{\dagger} \right).
\end{equation}
To maintain the nature of the new modes, it is essential that they preserve the bosonic commutation relations. This condition is satisfied when we choose the coefficients as
\begin{equation}
    \mu_{kq} = \frac{1}{2} \left( \sqrt{\frac{\omega_k }{\Omega_q}} + s_\mu\sqrt{\frac{\Omega_q}{\omega_k}} \right) \Lambda_{kq} 
\end{equation}
with $\mu=u,v$ and $s_u=-s_v=1$. The coefficients $\Lambda_{kq}$ satisfy the orthogonality condition  $\sum_q \Lambda_{kq} \Lambda_{k' q} = \delta_{kk'}$.} In \Cref{app:C} we show that we can obtain the coupling strengths $g_i$ and frequencies $\Omega_i$ in terms of the contributions to the original SD, $J(\omega) = \sum_i J^L_i (\omega)$ as 
\begin{equation}\label{eq:couplingstrengthRC}
    g_i^2 = \frac{1}{ \Omega_i } \int_0^{\infty} \textrm{d} \omega \, \omega J^L_i (\omega),\quad
    \Omega_i^2 = \frac{\int_0^{\infty} \textrm{d} \omega \, \omega J^L_i (\omega)}{\int_0^{\infty} \textrm{d} \omega \, \frac{J^L_i (\omega)}{\omega} }.
\end{equation}

\begin{figure*}
\centering
\subfloat[]{\label{fig:lorentzianmappedSD}{\includegraphics[width=0.5\textwidth]{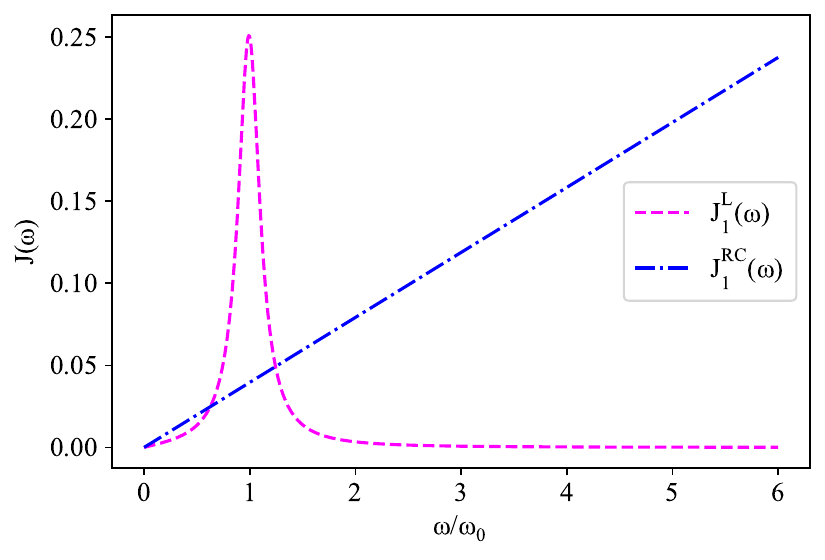}}}
\subfloat[]{\label{fig:oneRClorentzianSD}{\includegraphics[width=0.5\textwidth]{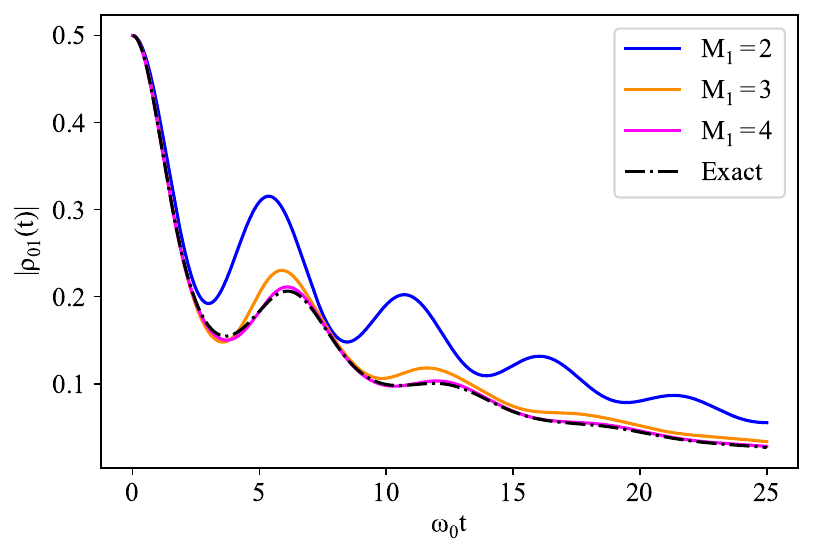}}}
\caption{The RC mapping applied to the pure dephasing model: Panel (a) shows a comparison between the original Lorentzian spectral density $J^L_1 (\omega)$, characterized by $\lambda_1 = \gamma_1 = 0.25$ and $\nu_1 = 1$, with the corresponding mapped spectral density $J^{RC}_1 (\omega)$, calculated using \Cref{eq:SDmapping}. The mapped spectral density exhibits the expected Ohmic behavior, demonstrating the linear dependence on frequency for $\omega/\omega_0\lesssim1$. Panel (b) shows the time evolution of the coherences in the pure dephasing model for the spectral density $J_0 (\omega) = J^L_1 (\omega)$, 
for the bath temperature $T = \omega_0/2$. The plot compares the results obtained using the exact solution with those from the RC mapping, where the dynamics are simulated with a single RC. The coherences are shown for different truncations of the RC dimension, $M_1$. The figure demonstrates that the RC mapping effectively captures the dynamics and closely approximates the exact solution, provided that $M_1$ is sufficiently large.}
\label{fig:subfigures}
\end{figure*}

The SD $J^{RC}_i (\omega) = \sum_k |h_{ik}|^2 \delta(\omega - \omega_k)$ characterises the interaction between the $i$-th RC and the residual environment. By analysing the Heisenberg equations of motion for the system and ensuring that the dynamics remain consistent before and after the mapping, we derive a relationship between $J^{RC}_i (\omega)$ and the $i$-th contribution to $J_{0} (\omega)$, $J^L_i (\omega)$, as detailed in \Cref{app:D}. This relationship is given by 
\begin{equation}\label{eq:SDmapping}
    J^{RC}_i (\omega) = \frac{2 g_i^2 J^L_i (\omega)}{\left( \mathcal{P} \int_{- \infty}^{\infty} \textrm{d} \omega' \,  \frac{J^L_i (\omega')}{(\omega' - \omega) }  \right)^2 + \pi^2 J^L_i (\omega)^2} \, ,
\end{equation}
where $\mathcal{P}$ denotes the Cauchy principal value. Ref.~\cite{10.1063/1.449017} demonstrated that the SD experienced by a particle coupled to an oscillator of frequency $\nu$, which subsequently interacts with an Ohmic bath, takes the form of a Lorentzian SD centered around $\nu$. An Ohmic spectral density is characterized by a linear dependence on frequency at low frequencies, typically described by $J(\omega) \propto \omega$ for small $\omega$. Thus, if the initial SD, $J^L_i (\omega)$, is Lorentzian, we expect that the mapped SD, $J^{RC}_i (\omega)$, will exhibit this linear, Ohmic behaviour. In \Cref{fig:lorentzianmappedSD}, the SD $J^L_1 (\omega)$, characterized by $\lambda_1 = \gamma_1 = 0.25\omega_0$ and $\nu_1 = \omega_0$ is displayed alongside its corresponding mapped SD $J^{RC}_1 (\omega)$, calculated using \Cref{eq:SDmapping}.

After performing the RC mapping, the next step is to solve the dynamics of the system. To this end, we treat the coupling between the two-level system and the RCs exactly while the coupling between the RCs and the residual environment is treated perturbatively, up to second order in the coupling strength. This approach allows us to rely on the standard Born-Markov approximation.
 {Within this regime and provided that the residual environment is in a thermal Gibbs state, the corresponding master equation in the Schr{\"o}dinger picture reads as
\begin{equation}\label{eq:RCME}
\begin{aligned}
    \frac{\partial \rho_{S'} (t)}{\partial t} &= - i [H_{S'}, \rho_{S'} (t)]  - \sum_{i \leq N, j, k} \left( \left[ A_i , A^i_{jk} \Gamma^+_i (\nu_{jk}) \rho_{S'} (t) \right] \right. \\
    & \left. + \left[\rho_{S'} (t) A^i_{jk} \Gamma^-_{jk} (\nu_{jk}), A_i \right] \right) \, ,
\end{aligned}
\end{equation}
where $\rho_{S'} (t)$ is the state of the enlarged system, $A_i = b_i + b_i^{\dagger}$, and $H_{S'} \ket{\phi_j} = \psi_j \ket{\phi_j}$, with $\nu_{jk} = \psi_j - \psi_k$. The operator $A^i_{jk}$ is defined as 
\begin{equation}
    A^i_{jk} = \bra{\phi_j} A_i \ket{\phi_k} \ket{\phi_j} \bra{\phi_k} \, .
\end{equation}
The functions $\Gamma^\pm_{i} (\omega)$ are given by 
\begin{equation}\label{eq:Gamma+}
    \Gamma^+_i (\omega) = \pi
    \begin{cases}
        J^{RC}_i (\omega) n_B (\omega),& \text{if } \omega > 0 \\
        \lim_{\omega \to 0} \left[ J^{RC}_i (\omega) n_B (\omega) \right],              & \text{if } \omega = 0 \\
        J^{RC}_i (|\omega|) \left( 1 + n_B (|\omega|) \right), & \text{if } \omega < 0
    \end{cases} \, ,
\end{equation}
\begin{equation}\label{eq:Gamma-}
    \Gamma^-_i (\omega) = \pi
    \begin{cases}
        J^{RC}_i (\omega) \left( 1 + n_B (\omega) \right) ,& \text{if } \omega > 0 \\
        \lim_{\omega \to 0} \left[ J^{RC}_i (\omega) \left( 1 + n_B (\omega) \right) \right],              & \text{if } \omega = 0 \\
         J^{RC}_i (|\omega|) n_B (|\omega|), & \text{if } \omega < 0
    \end{cases} \, .
\end{equation}}
For completeness, the derivation of these expressions is provided in \Cref{app:E}. We evolve the dynamics of the enlarged system starting from a given initial state. However, our interest lies in the dynamics of the two-level system alone, not the RCs. Therefore, we trace out the degrees of freedom associated with each RC, indexed by $i$, yielding the reduced density matrix of the two-level system, $\rho_S (t)$, as follows: 
\begin{equation}
    \rho_S (t) = \operatorname{tr}_{\bigotimes_{i=1}^N {RC}_i} \left( \rho_{S'} (t) \right) \, .
\end{equation}
Regarding the initial state of the enlarged system $\rho_{S'} (0)$, we assume that the two-level system and the RCs are prepared in a product state, i.e., $\rho_{S'} (0) = \rho_S (0) \bigotimes_{i=1}^N \rho^{RC}_i (0)$ where $\rho^{RC}_i (0)$ represents the initial state of the $i$-th RC which is thermalised to the residual environment. Thus, we write $\rho^{RC}_i (0) = {e^{- \beta \Omega_i b_i^{\dagger} b_i}}/{Z^{RC}_i}$, where $Z^{RC}_i = \operatorname{tr}_{RC_i} \left( e^{- \beta \Omega_i b_i^{\dagger} b_i} \right)$ is the partition function of the $i$-th RC. We are free to choose the initial state of the two-level system. 

To simulate the dynamics of the enlarged system, we must truncate its dimension. Specifically, we truncate the dimension of the $i$-th RC to include $M_i$ energy levels, ensuring that $M_i$ is large enough to guarantee that the results converge. To facilitate comparison with an exact solution, we consider the pure dephasing limit where the system operator $X = \sigma_z$. In this scenario, the general spin-boson model reduces to the exactly solvable pure dephasing model, as discussed in \Cref{sec:PDandAD}, with its solution provided in \Cref{eq:Exactlysolvablerhot}. \Cref{fig:oneRClorentzianSD} shows the time evolution of the coherences in the pure dephasing model obtained using both the exact solution and the RC mapping for the SD $J_0 (\omega) = J^L_1 (\omega)$, where $J^L_1 (\omega)$ is shown in \Cref{fig:lorentzianmappedSD}. As the SD consists of a single Lorentzian contribution, the dynamics are simulated using one RC. The temperature of the bath is $T = \omega_0/2$. The figure shows the evolution of the coherences for various truncations of the RC dimension, $M_1$. It is evident that the RC mapping accurately captures the dynamics of the pure dephasing model and closely approximates the exact solution, provided the RC dimension is sufficiently large.

Since the pure dephasing model can be simulated exactly, there is little practical value in using the RC mapping for this purpose. However, we use the pure dephasing model as a benchmark to determine the necessary dimensions of the RCs for specific parameter sets, including those related to the SD, the level-spacing of the two-level system, $\omega_0$, and the temperature of the bath, T. Once the appropriate dimensions are established, we apply them to the amplitude damping model, which is the primary focus of our study. This approach assumes that the dimensions sufficient for accurately capturing the dynamics of the pure dephasing model will also be adequate for the dissipative scenario. 

\subsection{Machine Learning Techniques}
\label{sec:MLtechniques}

\begin{figure*}
\centering
{\includegraphics[width=1.95\columnwidth]{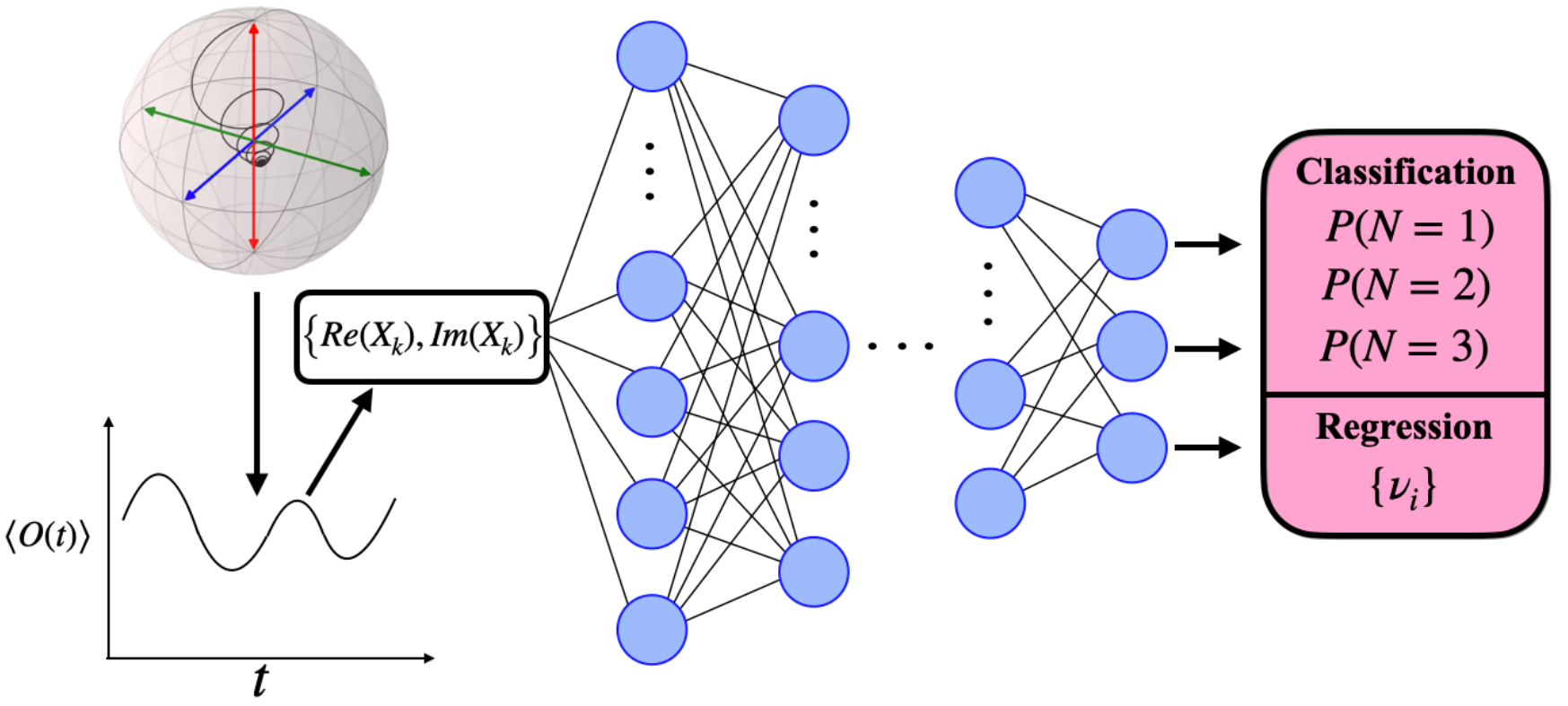}}
\caption{Sketch of the setup: Given the time evolution of an observable, denoted as $\langle O (t) \rangle$, we compute the corresponding Fourier coefficients $\{ X_k \}$. The real and imaginary parts of the coefficients, $\text{Re} (X_k)$ and $ \text{Im} (X_k)$, are then fed into a NN to infer properties of the SD characterising the system-environment interaction that produces the observed dynamics. For the classification task, the output layer contains three neurons, each representing the probability that the SD has one, two or three peaks. For the regression task, where we estimate the central frequency of the peaks, the output layer has a number of neurons matching the number of central frequencies to be predicted, or the number of peaks in the SD.}
\label{fig:sketchofprocedure}
\end{figure*}

In this work, we leverage ML techniques to characterise the SD that defines the system-environment interaction. Specifically, we employ an artificial NN composed of multiple artificial neurons, arranged in a series of  layers, as illustrated in~\Cref{fig:sketchofprocedure} \cite{Marquardt:2021,Marquardt:2023}.
Each neuron functions as a computational unit, receiving a set of inputs, denoted by ${x_i}$, and computing a weighted sum
\begin{equation}
z = \sum_i w_i x_i + b 
\end{equation}
with weights $w_i$ and bias  $b$. This result is then passed through a non-linear activation function $f$, producing the neuron's output $y = f(z)$. The activation function used in this work is the standard \emph{rectified linear unit} (ReLU) function, defined as 
\begin{equation}
f(z)=
\begin{cases}
z\quad\text{for}~z \geq 0,\\
0\quad\text{for}~z < 0.
\end{cases}
\end{equation}
Weights and biases within the network are free parameters that are optimized during the training process. The outputs generated by each layer serve as inputs for the subsequent layer. This means that as the input data propagates through the network, the outputs of deeper layers become increasingly complex functions of the initial input data. The first layer, known as the \emph{input layer}, receives the input data and passes it to the next layer without performing any computations. The final layer, referred to as the \emph{output layer}, computes the network's final output. The layers situated between the input and output layers are dubbed as \emph{hidden layers}. Note that we opt for this architecture due to its success in accomplishing the intended objective, without necessitating the use of a more complex architecture, such as a recurrent neural network \cite{Marquardt:2021}.

For the purpose of characterising the SD using a NN, we take as input the time evolution of a system observable, $\langle O (t) \rangle$. This time series may be obtained from experimental measurements in a laboratory, or, as in our case, generated by numerically simulating the system's dynamics. As each signal is a time series, we Fourier-decompose the signal. The Fourier coefficients are calculated as
\begin{equation}
\label{eq:Fourier1}
     X_k  = \sum_{p=0}^{P-1} \langle O (t_p) \rangle e^{- 2  \pi i k p /N } \, ,
\end{equation}
where $X_k \in\mathbb{C}$, $P$ represents the total number of time-steps, $\langle O (t_p) \rangle$ denotes the value at the $p^\text{th}$ sampled point, and $k=0,\dots,P-1$. The sum runs over all the sampled points in the time series. The original signal can then be reconstructed by inverting \Cref{eq:Fourier1}. Rather than using the time series directly, we split each Fourier coefficient into its real and imaginary parts, using such components as inputs to the NN. 

Using the resulting dataset, we begin by addressing the ternary classification task of identifying trajectories corresponding to SDs with one, two and three Lorentzian peaks. For this task, the output layer of the NN is configured with three artificial neurons, each of which compute weighted sums, denoted $z_j$, and apply the softmax activation function \cite{hastie2009elements}, given by
\begin{equation}
    f(z_j) = \frac{e^{z_j}}{\sum_{j=1}^{N_c} e^{z_j}} \, ,
\end{equation}
where $N_c$ represents the number of classes (here, $N_c=3$). The softmax function maps each output to a probability, indicating the likelihood that the input trajectory belongs to each class. For this classification task, we use the categorical cross-entropy as a loss function, a standard choice for classification problems. Given a dataset containing $N_t$ trajectories, we denote the true probability that the $i$-th trajectory belongs to the $j$-th class as $y_{ij}$, while $\hat{y}_{ij}$ denotes the NN's predicted probability for the same. The categorical cross-entropy loss function is then defined as \cite{murphy2012machine3}
\begin{equation}
    L(\hat{y}, y) = - \frac{1}{N_t} \sum_{i=1}^{N_t} \sum_{j=1}^{N_c} y_{ij} \log (\hat{y}_{ij}) \, .
\end{equation}

After categorising the dynamics based on the number of Lorentzian peaks in the SD, we proceed with a regression task to estimate the frequency of the peaks, {whose total number we assume to be known \emph{a priori}}. For this task, the output layer of the NN is configured with a number of artificial neurons matching the number of central frequencies to be predicted, depending on the number of peaks in the SD. We employ the linear activation function, defined as the identity function $f(z) = z$, which allows the network to generate unrestricted real-valued outputs \cite{geron2022hands, hastie2009elements}. In the regression tasks, we use the mean squared error (MSE) as the loss function. Let $N_{tN}$ represent the number of training examples containing $N$ peaks, $\nu_{ij}$ the true position of the $j$-th peak for the $i$-th example, and $\hat{\nu}_{ij}$ the corresponding predicted value. The MSE is then defined as~\cite{murphy2012machine3}
\begin{equation}\label{eq:meansquarederror}
L(\hat{\nu}, \nu) = \frac{1}{N_{tN} N} \sum_{i=1}^{N_{tN}} \sum_{j=1}^N (\nu_{ij} - \hat{\nu}_{ij})^2  .
\end{equation}
The task of training the networks reduces to an optimisation problem, where the aim is to find the set of network parameters that minimises the loss function. A schematic view of the setup is shown in \Cref{fig:sketchofprocedure}.

\section{Analysis and Results}
\label{sec:results}

In this section, we present the results of our numerical experiments, with classification outcomes discussed in~\Cref{sec:classification}, and regression results in~\Cref{sec:regression}. We focus on the case where $X = \sigma_x$ in \Cref{eq:H_I}, reducing the general spin-boson model to the amplitude damping model described in \Cref{sec:PDandAD}. For the initial state $\rho (0) = \ket{+}\bra{+}$, where $\ket{+} = \left( \ket{0} + \ket{1} \right)/ \sqrt{2}$, we generate a set of curves capturing the time evolution of $\langle \sigma_x (t) \rangle$. Each signal is sampled at $400$ points over the time interval $t \in [0, 50]$, ensuring a high resolution of the dynamics. As outlined in~\Cref{sec:MLtechniques}, instead of directly using the time series, we decompose each signal into $800$ real and imaginary parts of the Fourier coefficients $X_k$, which serve as inputs to the NN. For this reason, we build the input layers of the NNs with $800$ input neurons. The model architectures are optimised using the Keras tuner with Bayesian optimisation, and training is conducted using batch gradient descent with the Adam optimiser at a learning rate of $10^{-4}$. The code employed for data generation, the datasets, and the code used for subsequent analysis are available in the following GitHub repository \cite{github5}. 

We analyse trajectories corresponding to SDs that contain one, two or three Lorentzian peaks. For each Lorentzian component, we let $\gamma_i \in [0.15, 0.25]\omega_0$, $\lambda_i \in [0.1, 0.25]\omega_0$, and $\nu_i \in [0.54, 2]\omega_0$, while setting the environmental temperature to $T = \omega_0/2$. In order to determine the necessary dimensions of the RCs for different parameter configurations, we first use the pure dephasing model as a benchmark. By simulating the dynamics of this model using the RC mapping and comparing the results with the exact solution, we identify the minimum RC dimensions needed to achieve convergence within the RC mapping. For each Lorentzian peak in the SD, we assign an RC, with the corresponding dimensions shown in~\Cref{app:F}. Once the appropriate dimensions are identified, we apply the same RC dimensions to the amplitude damping model, which is the primary focus of our study. This approach assumes that the dimensions sufficient for accurately capturing the dynamics of the pure dephasing model will also be adequate for the dissipative scenario.

We generate $6\times10^3$ trajectories for each class - one, two, and three peaks- resulting in a total of $18\times10^3$ trajectories for classification, and further $6\times 10^3$ trajectories for each regression task. For each ML task, the dataset is divided into three subsets: $80\%$ of the data is allocated to training the model, $10\%$ is used as a validation set to monitor performance during training, and the remaining $10\%$ is reserved for the test set, which is used to assess the final performance of the network. 

\begin{figure}
\subfloat[]{\label{fig:trainingsetonepeak}{\includegraphics[width=1.05\columnwidth]{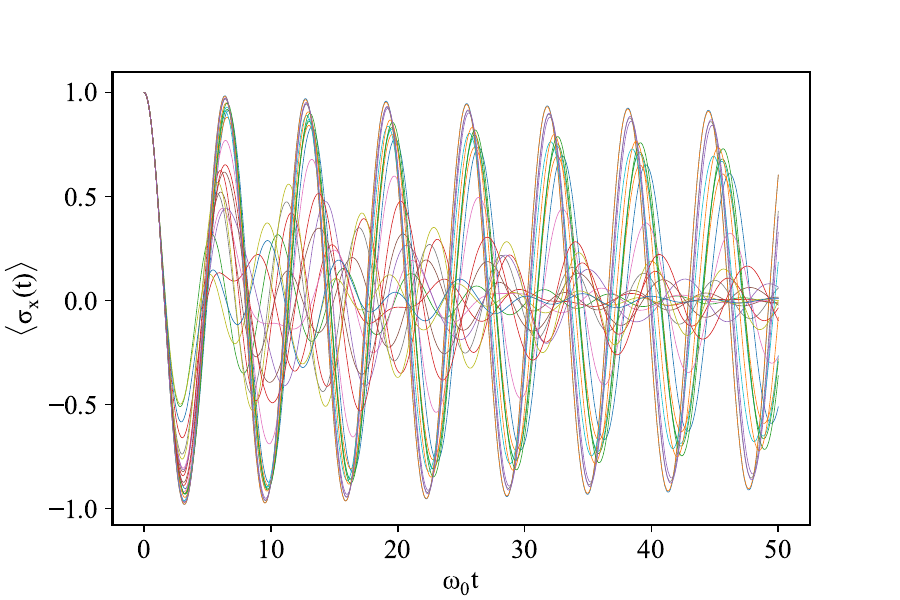}}}\\
\subfloat[]{\label{fig:trainingsettwopeaks}{\includegraphics[width=1.05\columnwidth]{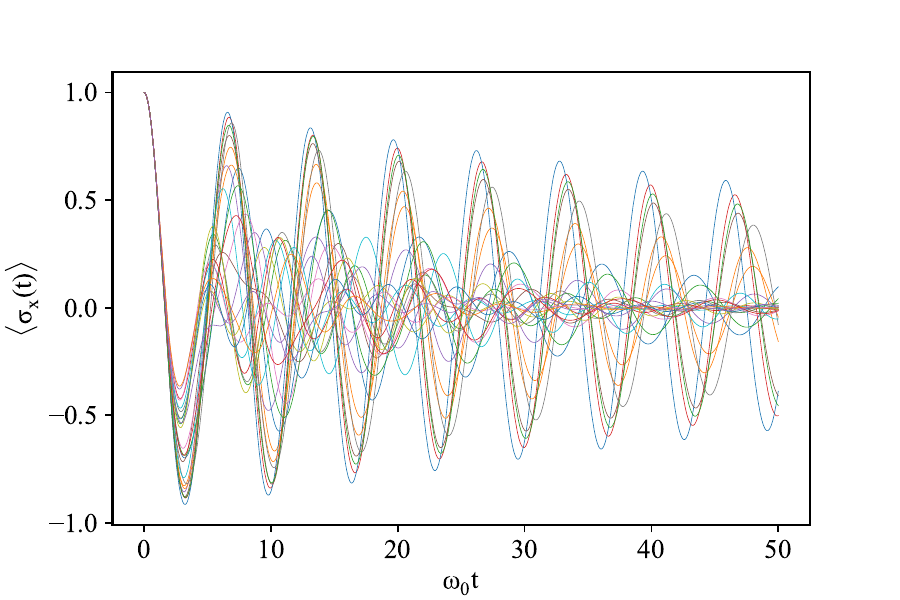}}}\\
\subfloat[]{\label{fig:trainingsetthreepeaks}{\includegraphics[width=1.05\columnwidth]{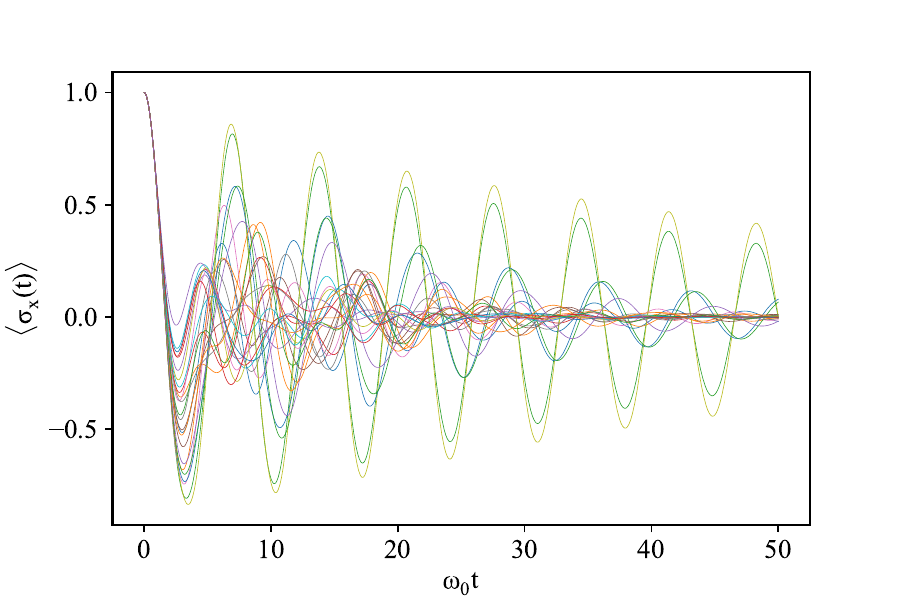}}}
\caption{Sample trajectories from the three classes of SDs: Panel (a) shows trajectories corresponding to SDs with one peak, panel (b) displays those with two peaks, and panel (c) illustrates three-peak trajectories, with $25$ samples in each plot. All trajectories were simulated using $\gamma_i \in [0.15, 0.25]\omega_0$, $\lambda_i \in [0.1, 0.25]\omega_0$, and $\nu_i \in [0.54, 2]\omega_0$, with $T = \omega_0/2$.}
\label{fig:trainingsets}
\end{figure}

\Cref{fig:trainingsets} shows 25 sample trajectories from each class. Specifically, \Cref{fig:trainingsetonepeak} displays trajectories with a single Lorentzian peak in their SD, \Cref{fig:trainingsettwopeaks} presents a subset of the trajectories with two peaks, and~\Cref{fig:trainingsetthreepeaks} illustrates those with three peaks. A clear trend of increased damping is observed as the number of peaks, and consequently, the number of RCs increases. In the one-peak trajectories, the oscillations maintain a relatively stable amplitude, fluctuating around zero, with many oscillatory maxima and minima reaching close to $\pm 1$. However, in the two-peak trajectories, the damping effect becomes more pronounced, resulting in a gradual reduction in amplitude over time. For the three-peak trajectories, the damping is strongest, producing small amplitudes that remain close to zero, with fluctuations that are considerably reduced compared to the one- and two-peak cases. This progression highlights how increasing the number of RCs in the SD amplifies the damping effect on the system's dynamics. 

\subsection{Classification of the Number of Peaks}
\label{sec:classification}

In this section, we present the results of our classification experiment, which aims to classify trajectories based on the number of peaks in their corresponding SD - specifically, identifying whether they have one, two, or three peaks. The optimal network configuration consists of 10 hidden layers, with the following neuron counts: $256$ neurons in the first and second layers, $96$ in the third, $224$ in the fourth, $32$ in the fifth, $128$ in the sixth, $448$ in the seventh, $64$ in the eighth, $320$ in the ninth, and $224$ in the tenth. The output layer contains $3$ neurons, each applying the softmax activation function. The optimal batch size for training is found to be $64$.

To evaluate the performance of the NN, we use the classification accuracy which is defined as the percentage of trajectories that are classified correctly. After $1000$ training iterations, the final training accuracy reached $99.04 \%$, while the test accuracy was $82.72\%$. When evaluating the test data, the model correctly labels $96.22 \%$ of the trajectories with one RC. For the trajectories with two RCs, the model achieves $75.53 \%$ accuracy, but mislabels $5.09 \%$ as having one RC and $19.38 \%$ as having three. For trajectories with three RCs, the model correctly classifies $75.64 \%$, with $0.34 \%$ mislabelled as having one RC, and $24.02 \%$ as having two. Notably, a substantial portion of the misclassifications occur between trajectories with two and three peaks.

\begin{figure}
\centering\includegraphics[width=0.5\textwidth]{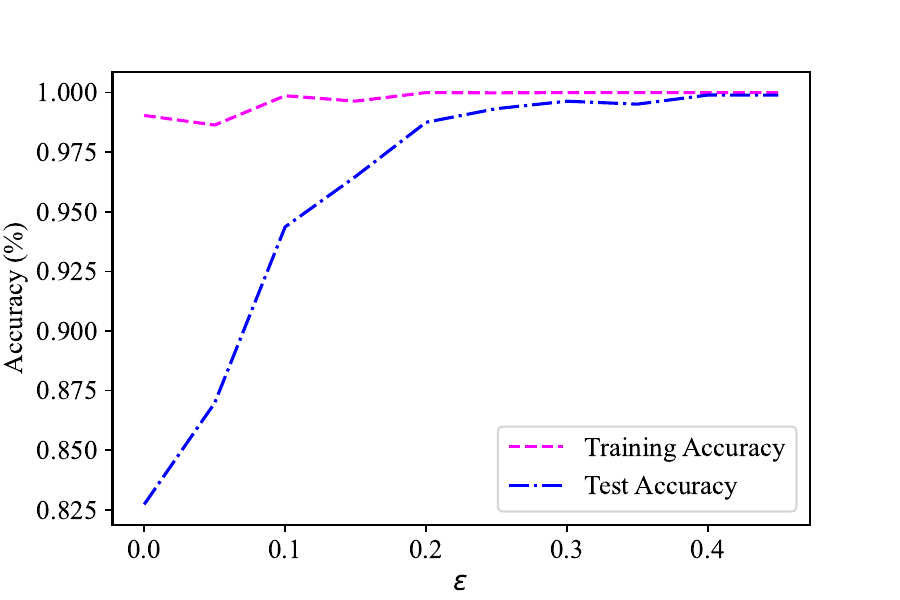}
	\caption{The training and test accuracy of the model as a function of the peak separation parameter $\varepsilon$ (in units of $\omega_0$). As $\varepsilon$ increases, the accuracy improves, demonstrating the model's enhanced ability to correctly classify trajectories with sufficiently separated peaks.} \label{fig:distancebetweenpeaksclassification}
\end{figure}

To address the misclassification issue between trajectories with two and three peaks, we introduce a condition that enforces a minimum separation between the centre frequencies of the Lorentzian peaks, denoted as $\varepsilon$. Specifically, we require that $ \omega_0|\nu_i - \nu_j| \geq \varepsilon $ for all ${i\neq j}=1,2,3$. For each value of 
$\varepsilon$, we filter the dataset to retain only those trajectories that meet these separation criteria. This approach allows us to train and test the model on progressively more selective datasets, ensuring that only trajectories with adequately separated peaks are included. We experiment with gradually increasing the value of $\varepsilon$ to determine the separation distance required for achieving a high classification accuracy. The training and test accuracies after $1000$ training iterations, as a function of $\varepsilon$, are shown in~\Cref{fig:distancebetweenpeaksclassification}. Our results indicate that as $\varepsilon$ increases, the model's accuracy improves, demonstrating that the model can effectively distinguish between trajectories with one, two, or three peaks in their SDs, provided that the peaks are sufficiently separated.

Following the implementation of the peak separation with varying $\varepsilon$, we aim to gain a deeper understanding of the factors that contribute to the model's challenges in classification, as well as those that enhance its performance. To this end, we examine the models performance for no enforced peak separation by analysing the correlation between the model's predicted classification probabilities and various trajectory features. For a trajectory characterised by a SD with $N$ peaks, we denote the model's predicted probability of assigning  exactly $N$ peaks as $P (N|N)$. To quantify the relation between the predicted probabilities and certain features, we use the Pearson correlation coefficient, which assesses the strength and direction of a linear relationship between two stochastic variables~\cite{wilcox2010fundamentals3, james2013introduction2}. A value of $1$ ($-1$) for the Pearson coefficient indicates perfect (anti-)correlation. Let $f_i$ denote the value of a specific feature for the $i^\text{th}$ trajectory in the set characterised by SDs with $N$ peaks, and $P_i (N|N)$ represent the predicted probability that the trajectory has $N$ peaks in its SD. The Pearson  coefficient between the feature and the probability is given by 
\begin{equation}
\label{eq:Pearson}
    C_{fM} \equiv \frac{ \sum_{i=1}^{N_t}  \Delta P_i (N|N)  \left( f_i - \overline{f} \right) }{\sqrt{\sum_{i=1}^{N_t}  \left( \Delta P_i (N|N) \right) ^2} \sqrt{\sum_{i=1}^{N_t} \left ( f_i - \overline{ f } \right )^2}} \, , 
\end{equation}
where $\overline{ P (N|N) }$ $\left(\overline{f}\right)$ is the average predicted probability (average value of the feature), $\Delta P_i (N|N) = P_i (N|N) - \overline{P(N|N)}$,  and $N_t$ is the total number of trajectories in the dataset. Using $C_{fM}$, we investigate how the average, maximum, and minimum values of the three parameters in the SD - the coupling constants, width parameters and central frequencies - affect the model's ability to correctly classify trajectories with two or three peaks. We also analyse the minimum distance between the peaks, and the average, maximum, and minimum distances of the peaks from the system's level spacing, $\omega_0$. The distance from $\omega_0$ is relevant because the system interacts more strongly with environmental frequencies close to $\omega_0$, which significant impacts the dynamics and, in turn, the model's ability to distinguish between different peak configurations. This correlation analysis, summarised in~\Cref{tab:correlation}, aims to provide deeper insights into the features that support or hinder the model's classification performance, thereby clarifying the challenges faced when peak separation is not enforced.
\begin{table}[]
    \centering
    \begin{tabular}{ccc}
    \hline
     \textbf{Feature}    &  ~~\textbf{Coefficient} $\mathbf{P ( 3 | 3)}$ & ~~\textbf{Coefficient} $\mathbf{P ( 2 | 2)}$ \\
     \hline\hline
     $\overline{\gamma}$ & $0.029$ & $- 0.080$ \\
     $\max \{ \gamma_i \}$ & $0.036$ & $- 0.057$\\
     $\min \{ \gamma_i \}$ & $- 0.013$ & $- 0.081$\\
     \hline
     $\overline{\lambda}$ & $0.467$ & $-0.266$ \\
     $\max \{\lambda_i\}$ & $0.400$ & $- 0.242 $ \\
     $\min \{ \lambda_i \}$ & $0.354$ & $- 0.226$ \\ 
     \hline
     $\overline \nu$ & $-0.013$ & $- 0.127$\\
     $\max \{\nu_i\}$ & $0.115$ & $- 0.062$ \\
     $\min \{ \nu_i \}$ & $- 0.115$ & $- 0.158$\\
     $\min \{ | \nu_i - \nu_j | \} $ & $0.369$ & $0.093$
     \\
     \hline
     $\overline{\{ |\omega_0 - \nu | \}}$ & $-0.012$ & $-0.168$ \\
     $\max \{ | \omega_0 - \nu_i | \}$ & $0.100$ & $-0.101$ \\
     $\min \{ | \omega_0 - \nu_i | \}$ & $-0.121$ & $-0.203$  \\
     \hline
    \end{tabular}
    \caption{Pearson correlation coefficients $P(2|2)$ and $P(3|3)$ between model-predicted classification probabilities and various features of the trajectories. The correlations are computed for the average, maximum, and minimum values of the coupling constants, width parameters, central frequencies, and the distances from the system's level spacing $\omega_0$, as well as the minimum distance between the peaks. These correlations highlight the features that influence the model's ability to accurately classify trajectories with two or three peaks.}
    \label{tab:correlation}
\end{table}

The width parameters $\gamma_i$ show weak correlations, with values near zero for both the two- and three-peak cases, indicating that the widths of the Lorentzians minimally influence the classification accuracy. In contrast, the coupling constants display strong positive correlations for three-peak classifications, particularly the average and maximum values, suggesting that higher coupling strengths enhance the model's ability to identify trajectories with three peaks. For two-peak classifications, however, the correlations with coupling strengths are consistently negative, implying that strong coupling makes it more challenging for the network to accurately identify two-peak trajectories.

The central frequencies generally show weaker correlations. For three-peak trajectories $\max \{ \nu_i\}$ has a weak positive correlation, while $\min \{ \nu_i\}$ shows a weak negative correlation, suggesting that greater peak spacing may improve classification accuracy. This is further supported by the positive correlation between the probability of correctly labelling a three-peak trajectory and the minimum distance between the peaks. In the two-peak case, $\min \{ \nu_i \}$ shows a weak negative correlation, and $\max \{ \nu_i \}$ shows a nearly negligible negative correlation. This suggests that the model's accuracy in identifying two-peak trajectories is less sensitive to peak spacing, as reflected in the minimal correlation between the minimum distance between peaks and the probability of correctly classifying a two-peak trajectory.

The last three rows of the table examine the correlation between the probability of correctly classifying two- and three-peak trajectories and the distance of the peaks' central frequencies from the system's level spacing, $\omega_0$. In both cases, the minimum distance from $\omega_0$ shows negative correlations, which is more pronounced for two-peaks trajectories. This suggests that the classification accuracy improves when at least one peak is close to $\omega_0$, likely due to stronger interactions between the system and the environment, which could enhance the network's ability to identify the characteristics of the trajectory.

In the three-peak case, the weak positive correlation between the the correct classification probability and the maximum distance from $\omega_0$ suggests having one peak is further from $\omega_0$ may make classification slightly easier. This could be because a well-separated is more distinctive, assisting the model in identifying a three-peak trajectory. Conversely, for two-peak trajectories, the weak negative correlation indicates that peaks further from $\omega_0$ make classification slightly harder, likely due to weaker system-environment interactions. Thus, for two-peak trajectories, as the peaks move further from $\omega_0$, classification becomes more challenging. This is consistent with the idea that stronger interactions near $\omega_0$ may enhance model accuracy.

\subsection{Regression of the central frequency of the peaks}
\label{sec:regression}

We now present the results of our regression experiment, which focuses on predicting the central frequency of the peaks in the SD after a trajectory has been classified as having one, two, or three peaks. 

The optimised NN architecture consists of $10$, 8 and 10 hidden layers, respectively, with neuron counts as in Table \ref{neuroncount}. 
\begin{table*}
\centering
\begin{tabular}{c|cccccccccc|cc}
\hline
\multirow{2}{*}{\textbf{Type of SD}} & \multicolumn{10}{c|}{\textbf{\bf $\#$ of neurons per hidden layer}}&{\textbf{MSE training set}} &{\textbf{MSE test set}} \\ \cline{2-11} 
& 1 & 2 & 3 & 4 & 5 & 6 & 7 & 8 &9 & 10 & ($\times10^{-6}$) & ($\times10^{-6}$)\\ \hline\hline
 \text{One-peak SD}& 224 & 288 & 128 & 224 & 32 & 288 & 32 & 480 & 512 & 352 & $6.4
 $ & $5.8$ \\
 \text{Two-peak SD}& 256 & 352 & 384 & 192 & 352 & 384 & 352 & 256 & &&$42.0$&$96.6$ \\
 \text{Three-peak SD}& 2512 & 288 & 128 & 32 & 64 & 32 & 448 & 512 & 320 & 224 &$182.4$ &$2999$\\
\hline
\end{tabular}
\caption{We report the neuron count per hidden layer of the optimized multi-layer NN architectures found for the trajectories stemming from considering a one-, two- and three-peak SD. Moreover, the last two columns of the table present the MSE, after $10^3$ training iterations, on training and test sets for each of the classes of SD considered.}
\label{neuroncount}
\end{table*}
As the goal is to predict the position of $k=1,2,3$ peak frequencies, the output layer of the optimal NN architectures has $k$ neurons, parameterized with the linear activation function. A batch size of $16$ is found to be optimal for all regression experiments.
\begin{figure*}
\centering
\subfloat[]{\label{fig:onepeakpredictedvreal}{\includegraphics[width=0.51\textwidth]{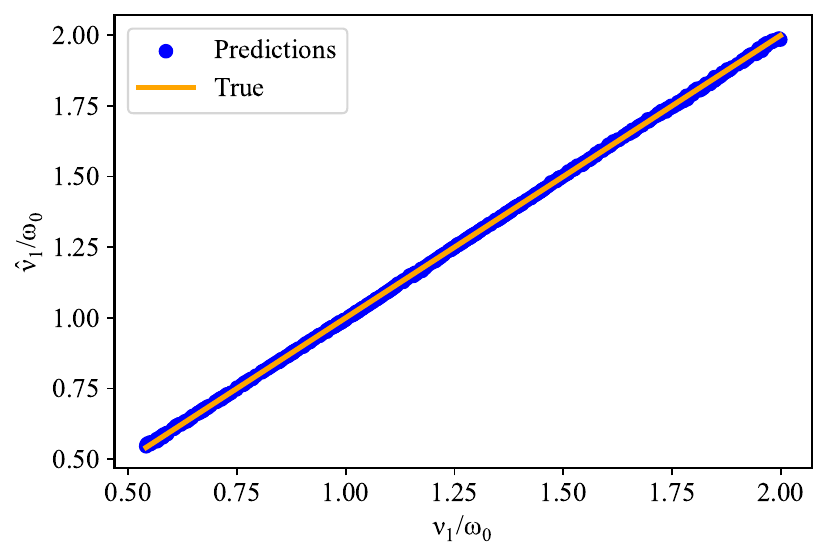}}}
\subfloat[]{\label{fig:onepeakbarchart}{\includegraphics[width=0.49\textwidth]{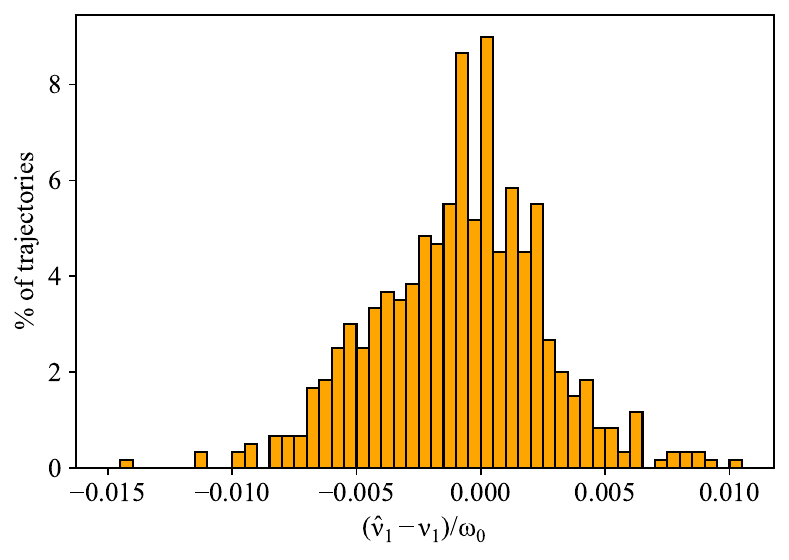}}}
\caption{Results of the regression analysis of the position of the peak in trajectories with a single-peak SD. Panel (a) shows the predicted peak position $\hat{\nu}_1$ against the true peak frequencies for the test set, along with a reference line indicating the ideal prediction $(\hat{\nu}_1 = \nu_1)$. Most points align closely with the reference line, reflecting high prediction accuracy. Panel (b) displays a bar chart showing the distribution of errors between predicted and actual values, with the majority of errors falling within small intervals, further confirming the model's strong performance. }
\label{fig:onepeakresults}
\end{figure*}

After $10^3$ training iterations, we evaluate the performance of the NNs on both the training and test sets, for each of the classes of SD being considered. The final values of the MSE 
for the training set and test set are reported in Table~\ref{neuroncount} as well: for a single-peak SD, we achieve very low values of the MSE for both training and test set,  indicating that the model is highly accurate in predicting the position of the peak. Notably, the test MSE is only slightly lower than the training MSE, suggesting effective generalization with minimal overfitting. This result highlights the model's robustness and its ability to maintain accuracy beyond the training data. For the two-peak SDs, the final MSEs grow by one order of magnitude compared to the one-peak case, given the increased complexity of the problem being faced. The trend is further reinforced when addressing trajectories stemming from three-peak SDs, whose corresponding MSEs are significantly larger than in the previous cases. 

\begin{figure*}
\centering
\subfloat[]{\label{fig:twopeakspredictednu1vreal}{\includegraphics[width=0.502\textwidth]{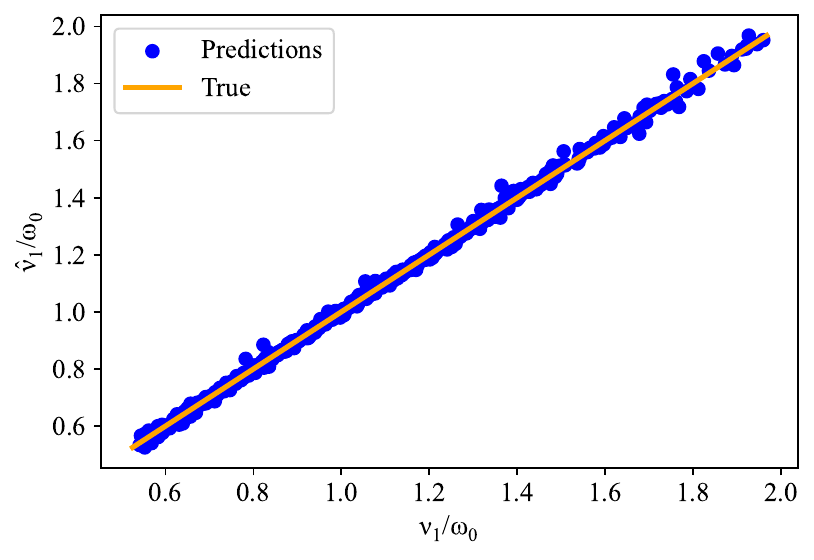}}}
\subfloat[]{\label{fig:twopeaksbarchartnu1}{\includegraphics[width=0.498\textwidth]{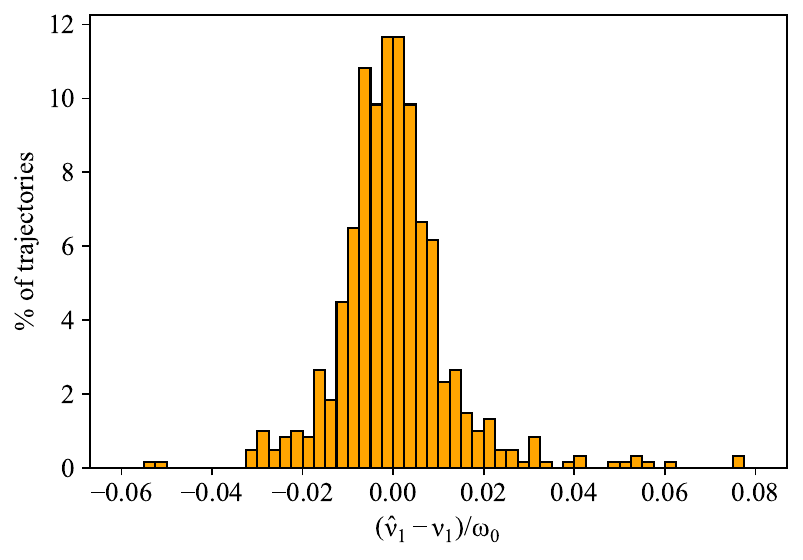}}}
\hfill
\subfloat[]{\label{fig:twopeakspredictednu2vreal}{\includegraphics[width=0.501\textwidth]{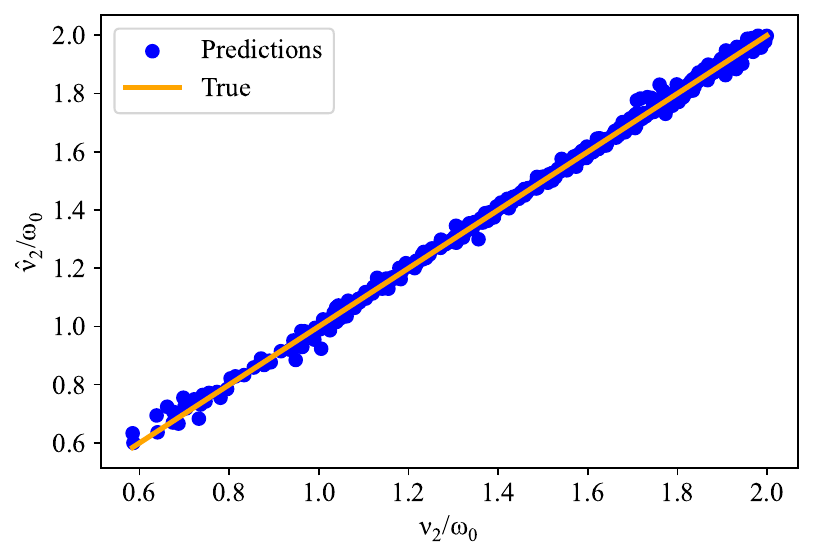}}}
\subfloat[]{\label{fig:twopeaksbarchartnu2}{\includegraphics[width=0.499\textwidth]{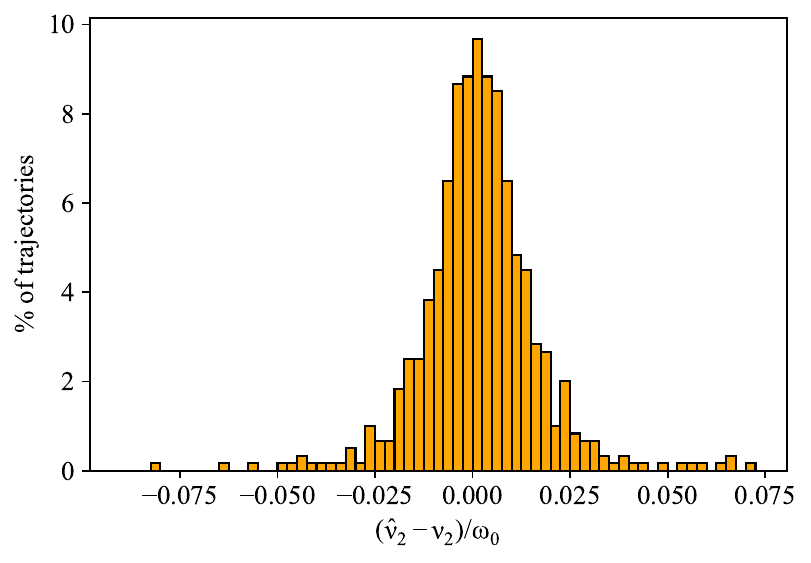}}}
\caption{Results of the regression analysis of peak frequency in trajectories with two peaks: Panel (a) shows the predicted peak position, $\hat{\nu}_1$, against the true peak position for the test set, with a reference line representing the ideal prediction $\hat{\nu}_1 = \nu_1$. Panel (b) presents a bar chart showing the distribution of errors between the predicted and true values for $\hat{\nu}_1$. Similarly, Panel (c) displays the predicted peak position $\hat{\nu}_2$ against the true peak position for the test set, along with a reference line indicating the ideal prediction $\hat{\nu}_2 = \nu_2$. Panel (d) displays a bar chart depicting the error between the predicted and true values for $\hat{\nu}_2$.}
\label{fig:twopeakresults}
\end{figure*}

\begin{figure*}
\centering
\subfloat[]{\label{fig:threepeakspredictednu1vreal}{\includegraphics[width=0.501\textwidth]{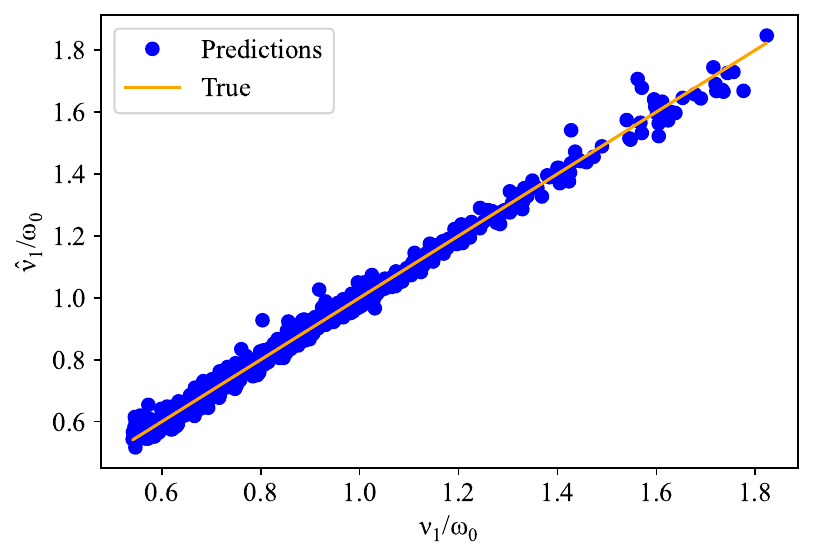}}}
\subfloat[]{\label{fig:threepeaksbarchartnu1}{\includegraphics[width=0.499\textwidth]{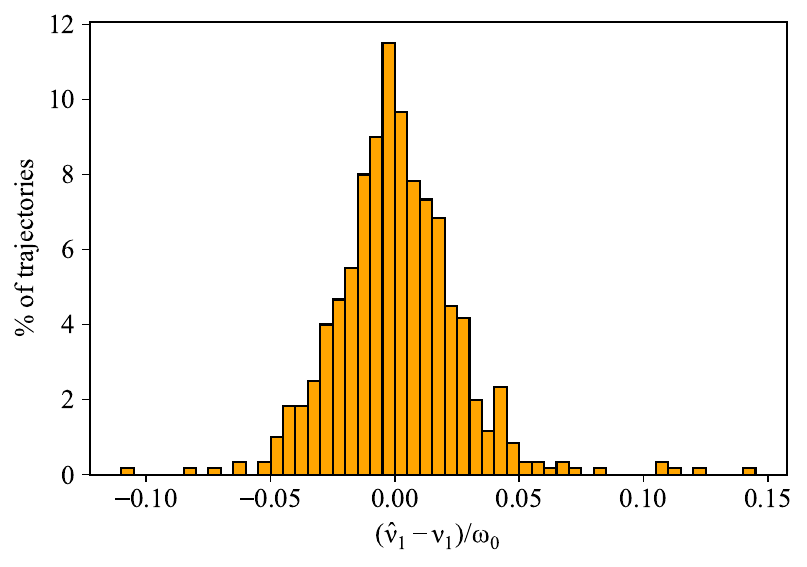}}}
\hfill
\subfloat[]{\label{fig:threepeakspredictednu2vreal}{\includegraphics[width=0.501\textwidth]{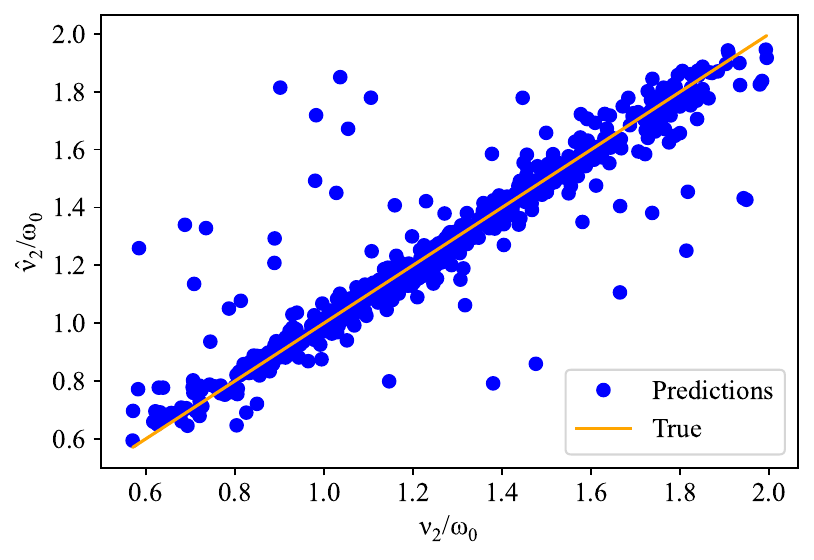}}}
\subfloat[]{\label{fig:threepeaksbarchartnu2}{\includegraphics[width=0.499\textwidth]{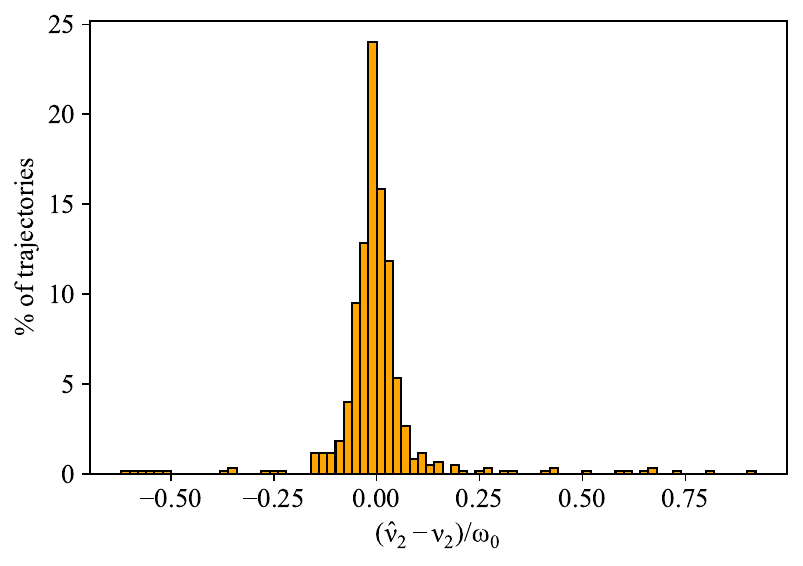}}}
\hfill
\subfloat[]{\label{fig:threepeakspredictednu3vreal}{\includegraphics[width=0.501\textwidth]{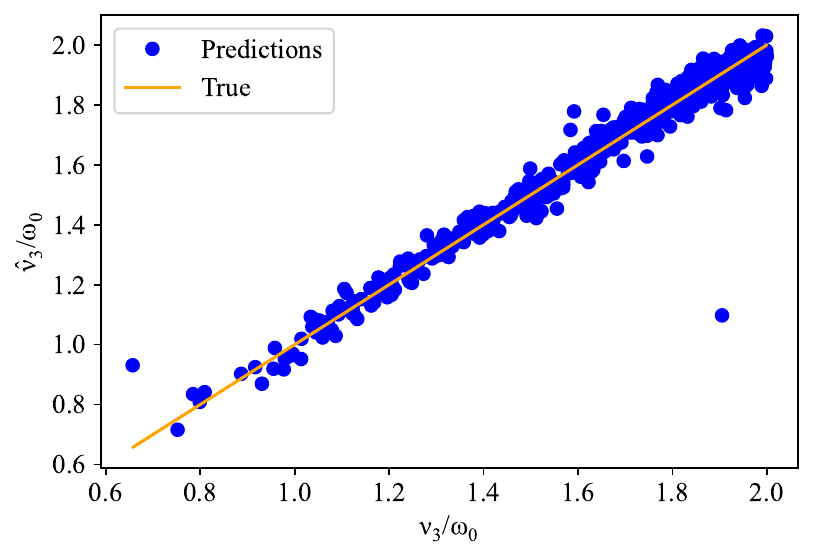}}}
\subfloat[]{\label{fig:threepeaksbarchartnu3}{\includegraphics[width=0.499\textwidth]{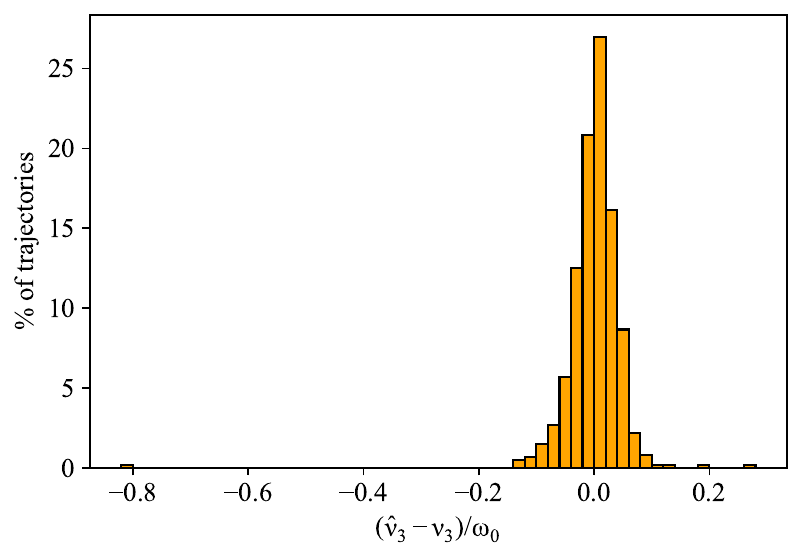}}}
\caption{Results of the regression analysis for peak frequency in trajectories with three peaks: Panels (a), (c), and (e) show the predicted peak frequency $\hat{\nu}_1$, $\hat{\nu_2}$, and $\hat{\nu_3}$, respectively, plotted against the true peak frequency for the test set. Each plot includes a reference line representing the ideal predictions, where $\hat{\nu}_i = \nu_i$. Panels (b), (d), and (f) present bar charts depicting the distribution of errors between the predicted and true values for $\hat{\nu}_1$, $\hat{\nu}_2$ and $\hat{\nu}_3$, respectively. }
\label{fig:threepeakresults}
\end{figure*}

In order to dig deeper into the quality of such findings, in Figs.~\ref{fig:onepeakpredictedvreal}-to-\ref{fig:threepeaksbarchartnu3} we assess the model's performance by comparing the predicted peak central frequency with the true values and analyzing the error distribution across the test set.  Figs.~\ref{fig:onepeakpredictedvreal}, \ref{fig:twopeakspredictednu1vreal}, and \ref{fig:twopeakspredictednu2vreal} show that the predicted values of peak frequencies are in strong agreement with the actual values when one- and two-peak SDs are considered. The case of three peaks, instead, while providing a very close alignment between predictions and actual value of $\nu_1$ and $\nu_3$, reveals a broader spread around the diagonal for $\nu_{2}$, suggesting that the model struggles more to accurately predicting the value of such peak frequency.  
However, the study of the distribution of errors associated with the trajectory-dependent predictions reported in the bar charts of Figures ~\ref{fig:onepeakbarchart}, \ref{fig:twopeaksbarchartnu1}, \ref{fig:twopeaksbarchartnu2}, \ref{fig:threepeaksbarchartnu1}, \ref{fig:threepeaksbarchartnu2}, and \ref{fig:threepeaksbarchartnu3}, shows that most trajectories exhibit errors close to zero, with only a small portion showing larger deviations, thus indicating that the model's performance is robust across the test set. 

Overall, these results demonstrate that the NN is highly effective in predicting the frequency of a single peak in the SD. The low error values and strong agreement between predicted and actual values indicate that the model can generalise well to unseen data.

\begin{figure}
\centering\includegraphics[width=\linewidth]{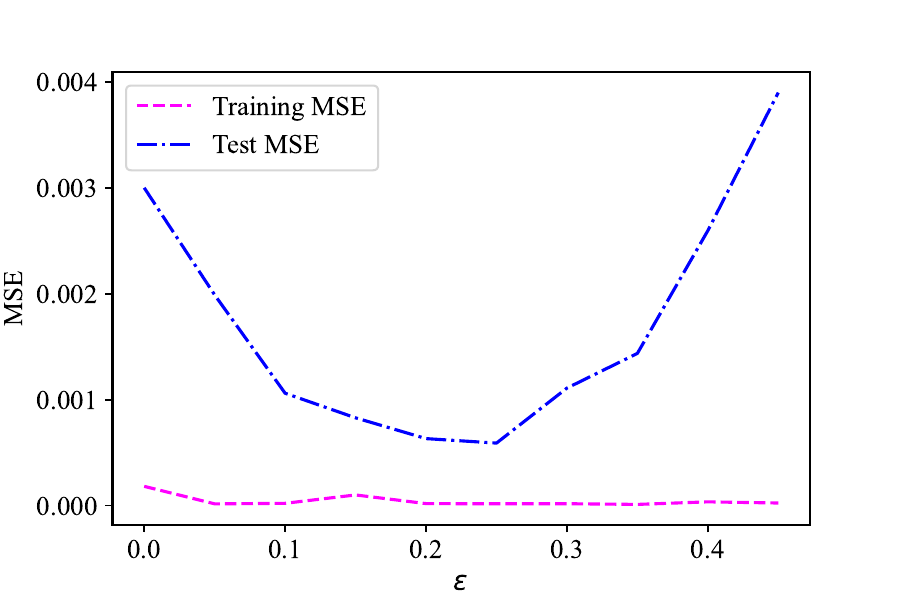}
	\caption{MSE evaluated on the training and test sets as a function of the minimum peak separation distance, $\varepsilon$ (in units of $\omega_0$), for the regression of three peak frequency. } \label{fig:distancebetweenpeaksregression}
\end{figure}

Following the approach used in the classification task to mitigate the model's confusion between two- and three-peak trajectories, we introduce a condition requiring the peaks to be separated by a minimum distance $\varepsilon$. Specifically, we impose the previously used constraint $\omega_0|\nu_i - \nu_j| \geq \varepsilon~\forall{i\neq j}=1,2,3$ which sets a minimum separation between the peaks. By progressively increasing $\varepsilon$, we aim to investigate how this enforced separation influences the model's regression performance. The training and test MSEs after $10^3$ training iterations, corresponding to different values of $\varepsilon$, are presented in~\Cref{fig:distancebetweenpeaksregression}.

When no minimum separation is enforced, the MSE for both datasets is relatively high, though the test MSE reaches its maximum at the largest values of $\varepsilon$. The MSE on the training set remains consistently low across the range of values of $\varepsilon$ being considered. As the peak separation increases to $0.25$, the MSE evaluated on the test set decreases steadily, suggesting that the enforced separation simplifies the task, allowing the model to generalize better within this range. For $\varepsilon>0.25$, however, the test set MSE begins to increase, indicating that further increasing the separation does not yield additional improvements in prediction accuracy. It is important to note that dataset sizes decrease for higher values of $\varepsilon$ due to filtering, which may contribute to this rise in test error. Expanding the dataset for larger values of $\varepsilon$ could potentially improve performance.

\begin{figure*}
\centering
\subfloat[]{\label{fig:threepeakspredictednu1vreal_aftersplitting}{\includegraphics[width=0.51\textwidth]{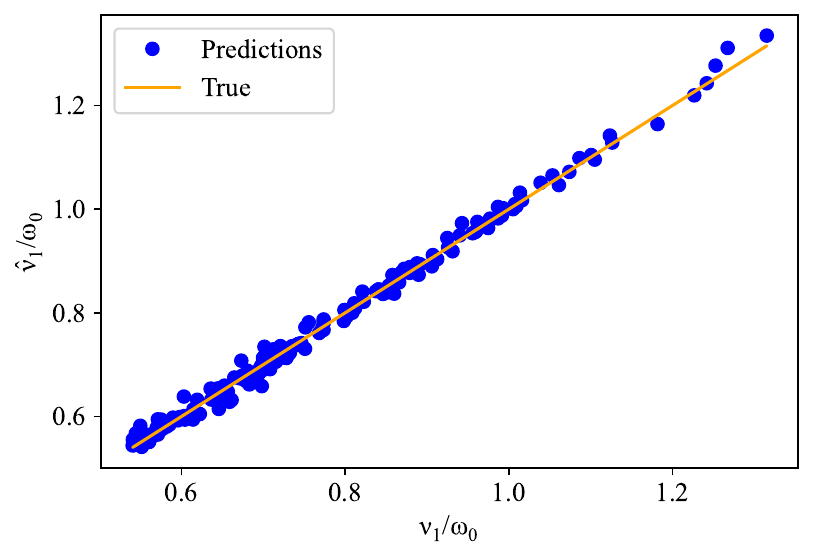}}}
\subfloat[]{\label{fig:threepeaksbarchartnu1_aftersplitting}{\includegraphics[width=0.49\textwidth]{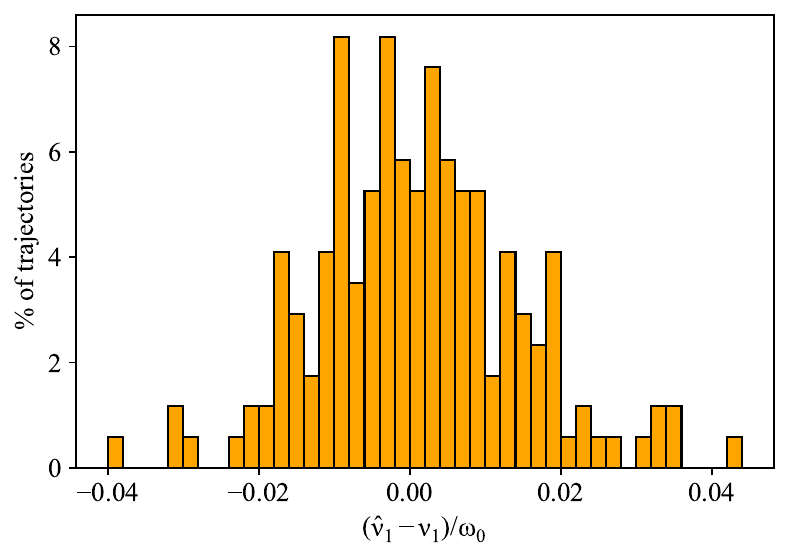}}}
\hfill
\subfloat[]{\label{fig:threepeakspredictednu2vreal_aftersplitting}{\includegraphics[width=0.51\textwidth]{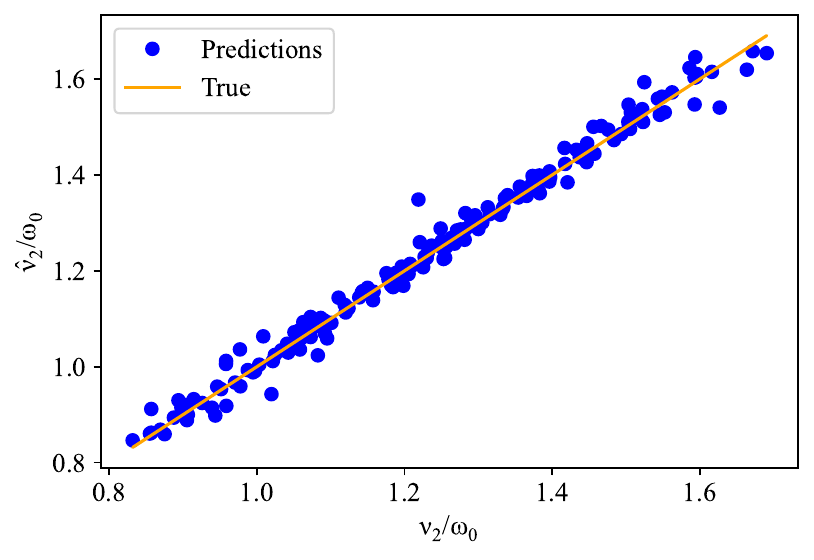}}}
\subfloat[]{\label{fig:threepeaksbarchartnu2_aftersplitting}{\includegraphics[width=0.49\textwidth]{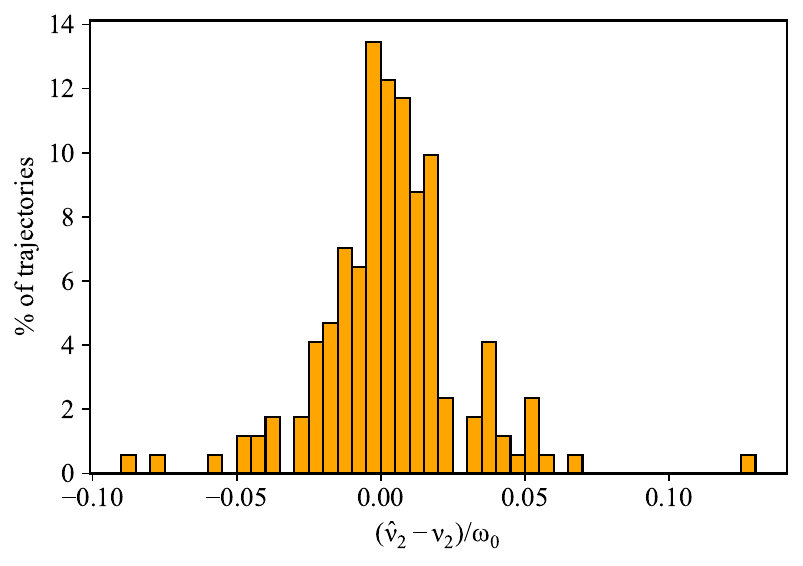}}}
\hfill
\subfloat[]{\label{fig:threepeakspredictednu3vreal_aftersplitting}{\includegraphics[width=0.51\textwidth]{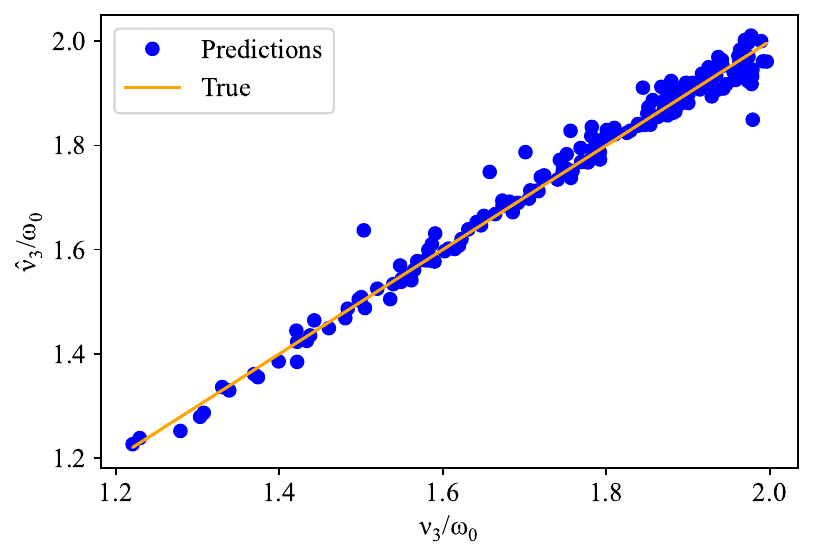}}}
\subfloat[]{\label{fig:threepeaksbarchartnu3_aftersplitting}{\includegraphics[width=0.49\textwidth]{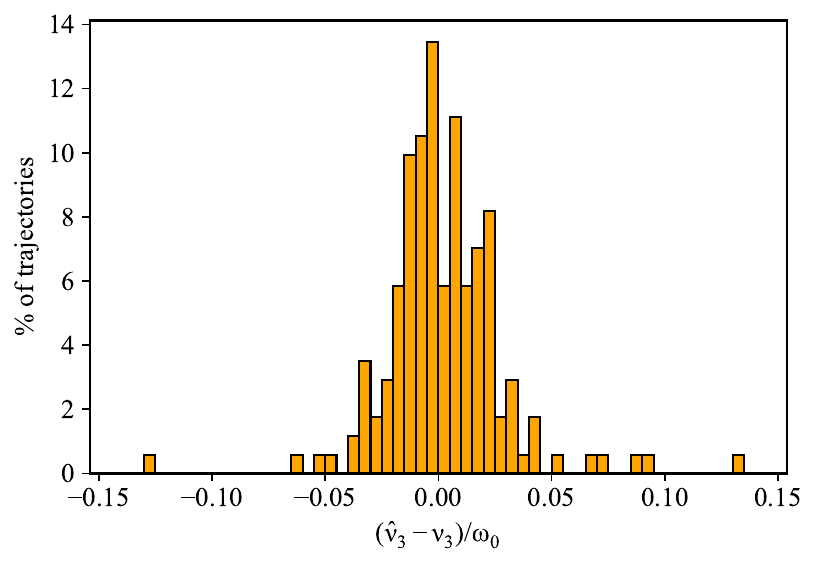}}}
\caption{Results of the regression analysis for peak frequency in trajectories with three peaks, with a minimum separation of $\varepsilon = 0.25$: Panels (a), (c), and (e) display the predicted peak frequency $\hat{\nu}_1$, $\hat{\nu_2}$, and $\hat{\nu_3}$, respectively, plotted against the true peak frequency for the test set, with a reference line indicating ideal predictions, where $\hat{\nu}_i = \nu_i$. Panels (b), (d), and (f) present bar charts showing the error distribution between predicted and true values for $\hat{\nu}_1$, $\hat{\nu}_2$ and $\hat{\nu}_3$, respectively.}
\label{fig:threepeakresults_aftersplitting}
\end{figure*}

To further investigate the impact of enforcing a minimum separation of $\varepsilon = 0.25$ between peaks, we analyse the relationship between the predicted and true peak frequency, as shown in~\Cref{fig:threepeakspredictednu1vreal_aftersplitting,fig:threepeakspredictednu2vreal,fig:threepeakspredictednu3vreal}. The corresponding error distributions for $\nu_1$, $\nu_2$, and $\nu_3$, are presented in Figures \ref{fig:threepeaksbarchartnu1_aftersplitting}, \ref{fig:threepeaksbarchartnu2_aftersplitting}, and \ref{fig:threepeaksbarchartnu3_aftersplitting}. For $\nu_1$, the scatter plot before the forced separation already showed a strong alignment between the predicted and true values along the diagonal, indicating that the model performed well, with only minor deviations. After the forced separation, the alignment remains tight, and predictions continue to be consistent. While the forced separation hasn't markedly changed the model's performance for $\nu_1$, it has led to a slight refinement in the error distribution, suggesting improved stability without sacrificing accuracy.

For $\nu_2$, the model initially encountered greater challenges. Before the forced separation, the scatter plot showed a broader spread and larger deviations from the diagonal, indicating difficulties in accurately predicting $\nu_2$. With the enforced minimum separation, the spread in predictions has reduced significantly, though some deviations from the diagonal are still visible. While $\nu_2$ remains more challenging to predict than $\nu_1$, the forced separation has clearly improved prediction accuracy, reducing the overall magnitude of errors.

For $\nu_3$, the scatter plot before forced separation showed some noticeable deviations from the ideal line, indicating that $\nu_3$ was more difficult to predict than $\nu_1$, though less challenging than $\nu_2$. After enforcing the minimum separation, the scatter plot shows noticeable improvement, with predictions being more tightly aligned along the diagonal, suggesting enhanced accuracy for $\nu_3$. Additionally, the error distribution for $\nu_3$ is now more concentrated around zero, reflecting a tighter overall distribution compared to the results before separation. While some deviations remain, the forced separation has clearly contributed to improved prediction accuracy.

In summary, the forced separation has led to improvements across all three peaks. For $\nu_1$, the model continues to perform well, with a refined error distribution. The most pronounced improvements post-separation are seen in $\nu_2$ and $\nu_3$, both of which show significant enhancements in alignment and error distribution. Although  $\nu_2$ and $\nu_3$ remain more challenging for the model than $\nu_1$, the benefits of the separation are evident.

\section{Conclusions}
\label{sec:conclusions}

In this study, we have demonstrated that, given the time evolution of a system observable, a NN can classify the SD as having one, two or three Lorentzian peaks in standard open quantum system scenarios. Beyond classification, we have also shown that a NN can effectively perform regression to estimate the frequency of these peaks. Leveraging the RC mapping allowed us to characterise SDs that lead to long-lived correlations and hinder the derivation of a master equation within the Born-Markov regime. Thus, we have demonstrated that the combination of the RC mapping and ML techniques allows us to characterise the SD beyond the weak coupling regime, where traditional methods may falter. 

To gain insights into the factors that contribute to high classification accuracy, as well as those that introduce challenges, we have analysed correlations between the predicted classification probabilities and specific features of the trajectories. These features include the values of the parameters in the SD, peak spacing, and the distances of the peaks from the bare oscillator frequency. Additionally, we introduced a forced separation between peaks and demonstrated that increasing this separation enhances both classification and regression accuracy.

Looking ahead, the techniques and insights developed in this paper open several promising avenues for further research. For instance, we could explore the application of this approach to more complex quantum systems, such as multi-qubit or multi-level systems. Additionally, expanding our analysis to include a wider variety of SD classes beyond those considered in this paper would allow for a more comprehensive understanding of environmental influences on quantum systems. Importantly, we would like to emphasise that the methods presented in this work do not depend on how the information about the system's dynamics is obtained. {On one hand, one can rely on different methods to simulate the dynamics, such as Hierarchical Equation Of Motion (HEOM)~\cite{Tanimura:1989,Tanimura:1990}, Time-Evolving Matrix Product Operators (TEMPO)~\cite{Lovett:2018}, or Time-Evolving Density with Orthogonal Polynomials Algorithm (TEDOPA)~\cite{Chin:2010,Prior:2010,Tamascelli:2019,Nusseler:2022}. On the other hand, it would be interesting to go beyond synthetic data, considering experimental sets instead}. We expect that these methods will perform equally well with the latter, opening the door to practical applications in real-world quantum systems. 

The methodology developed in this paper, along with the case studies analysed therein, demonstrate the effectiveness of ML techniques for characterising environments with arbitrary SDs. This approach provides a reliable tool for environment characterisation, which is essential for advancing quantum control and error correction strategies.

\acknowledgements
 JB and MP thank the Leverhulme Trust Doctoral Scholarship grant LINAS. We acknowledge support from
the European Union's Horizon Europe EIC-Pathfinder
project QuCoM (101046973), the Royal Society Wolfson Fellowship (RSWF/R3/183013), the UK EPSRC (grants EP/T028424/1), the Department for the Economy of Northern Ireland under the US-Ireland R\&D Partnership Programme, the "Italian National Quantum Science and Technology Institute (NQSTI)" (PE0000023) - SPOKE 2 through project ASpEQCt, the "National Centre for HPC, Big Data and Quantum Computing (HPC)" (CN00000013) - SPOKE 10 through project HyQELM.

\appendix

\section{The Correlation Function of a Bosonic Bath}
\label{app:A} 

In this Appendix, we explicitly derive the correlation function for an arbitrary quantum system interacting with an environment consisting of infinitely many independent harmonic oscillators, as shown in \Cref{eq:bosoniccorrfunc} of the main text. Given an interaction Hamiltonian in the form of \Cref{eq:H_I} and the bath operator $B$ as defined by \Cref{eq:bath_operator}, we can compute the correlation function, which is defined as
\begin{equation}
     C (\tau) = \langle \tilde{B} (\tau) \tilde{B} (0) \rangle_E = \operatorname{tr}_E \left( \tilde{B} (\tau) \tilde{B} (0) \rho_E \right) \, .
     \label{eq:corr}
\end{equation}
We move to the interaction picture using the relation $\tilde{B}(\tau) = e^{i \tau H_E} B e^{-i \tau H_E}$, where $H_E = \sum_k \omega_k b_k^{\dagger} b_k$ represents the Hamiltonian for a collection of independent harmonic oscillators. Consequently, we have
\begin{align}\label{eq:B_int_pic}
    B (\tau) & = \sum_k \left( t_k b_k^{\dagger} e^{i \omega_k \tau} + t_k^* b_k e^{-i \omega_k \tau} \right).
\end{align}
If we assume that the environment is in thermal equilibrium at a temperature $T$, then $\rho_E$ is represented by a thermal Gibbs state of the form $\rho_E = {e^{- \beta H_E }}/{\mathcal{Z}_E}$,
where $\mathcal{Z}_E$ is the reservoir partition function, the expression for the correlation function reads 
\begin{equation}
     C (\tau) = \sum_k | t_k |^2 \left(  \langle b_k^{\dagger} b_k \rangle_E \, e^{i \omega_k \tau} + \langle b_k b_k^{\dagger}  \rangle_E  \,e^{-i \omega_k \tau}  \right),
\end{equation}
where we used $\langle b_k b_l \rangle_E = \langle b_k^{\dagger} b_l^{\dagger} \rangle_E = 0$ and the fact that $\langle b_k b_l^{\dagger} \rangle_E$ and $\langle b_k^{\dagger} b_l \rangle_E $ are non-zero if and only if $k = l$. In addition, we note that the quantity $\langle b_k^{\dagger} b_k \rangle_E = N_k = (e^{\beta \omega_k} - 1)^{-1}$ is the mean occupation number of the $k$-th mode of the environment. Finally, assuming that the bath modes form a continuum, we obtain \Cref{eq:bosoniccorrfunc} in the main text. 

\section{The Pure Dephasing Model}
\label{app:B}

We now derive the equations that describe the dynamics of the pure dephasing model introduced in \Cref{sec:PDandAD}. Working in the interaction picture, we being by deriving an expression for the unitary evolution operator $\tilde{U} (t)$ which governs the dynamics of the composite system. It is important to note that the two-time commutator of the interaction Hamiltonian is non-zero, and is given by
\begin{equation}
\label{eq:SB_commutator}
    \left[ \tilde{H}_I (t) , \tilde{H}_I (t') \right] = - 2 i \, \mathbb{1}_{S} \otimes \sum_k |t_k|^2 \sin (\omega_k (t - t')) ,
\end{equation}
where $\mathbb{1}_S$ represents the identity operator acting on the system only. This commutator is useful for evaluating the time evolution operator, which is expressed as
\begin{equation}
    \tilde{U} (t) = \mathcal{T}_{\leftarrow} \exp\left[ -i \int_0^t \textrm{d} \tau \, \tilde{H}_I (\tau) \right] \, ,
\end{equation}
where $\mathcal{T}_{\leftarrow}$ denotes the time-ordering operator. Following the approach outlined in Ref. \cite{Lidar:2001} (and further discussed in Ref.~\cite{Schaller:2014}), we can formally discretise the integral in the exponent of the unitary evolution operator. We denote the discrete contribution at each time step as $\mathcal{H}_n = - i \tilde{H}_I (n \textrm{d}t)$, where $\textrm{d}t = t/N$ and $N$ represents the total number of time intervals. In the limit as $N \to \infty$, we obtain
\begin{equation}
    \tilde{U} (t) = \mathcal{T}_{\leftarrow} \lim_{\textrm{d}t \to 0} \exp\left[ \sum_{n = 1}^N \mathcal{H}_n\, \textrm{d}t\right] \, .
\end{equation}
We use a generalisation of the Baker-Campbell-Hausdorff formula to evaluate the exponential
\begin{equation}
    e^{ \sum_{n=1}^N \mathcal{H}_n}  = \left( \prod_{n = 1}^N e^{ \mathcal{H}_n} \right) \left( \prod_{n < m } e^{ -  \frac{1}{2} \left[ \mathcal{H}_n, \mathcal{H}_m \right]} \right) \, ,
\end{equation}
which is valid under the condition that the second order commutators vanish. Given that the commutator in the first exponent is just a complex number, we can omit the time ordering operator, leading to
\begin{equation}
    \tilde{U} (t) = \lim_{\textrm{d}t \to 0 } \prod_{n < m } e^{  - \frac{1}{2} \left[  \mathcal{H}_n, \mathcal{H}_m \right] (\textrm{d}t)^2} \prod_n e^{ \mathcal{H}_n \textrm{d} t} \, .
\end{equation}
Recombining the exponentials of the operators we find
\begin{equation}
\begin{aligned}
    \tilde{U} (t) & = \lim_{\textrm{d}t \to 0} e^{ - \frac{1}{2} \sum_{n<m} \left[ \mathcal{H}_n, \mathcal{H}_m \right] (\textrm{d}t)^2} e^{ \sum_n \mathcal{H}_n \textrm{d} t }\\
    & = e^{ \frac{1}{2} \int_0^t \textrm{d}t_1 \int_0^{t_1} \textrm{d} t_2 \left[ \hat{H}_I (t_2), \hat{H}_I (t_1) \right]} e^{ -i \int_0^t \textrm{d} \tau \hat{H}_I (\tau) },
\end{aligned}
\end{equation}
where the first exponent, due to the commutation in \Cref{eq:SB_commutator}, only introduces a global phase. Consequently, the dynamics of the system are governed entirely by the operator
\begin{equation}
    e^{-i \int_0^t \tilde{H}_I (\tau) \textrm{d} \tau } = e^{{\sigma}_z \otimes\sum_k \left( \alpha_k (t) b_k^{\dagger} - \alpha_k^* (t) b_k \right)} \equiv e^{\sigma_z \otimes A(t)} \, ,
\end{equation}
with $\alpha_k(t) = t_k \left( 1 - e^{i \omega_k t}\right)/\omega_k$. It is convenient to expand this operator as
\begin{align}
    e^{\sigma_z \otimes A(t)} & = I \otimes \sum_{n=0}^{\infty} \frac{A(t)^{2n}}{2n!} + \sigma_z \otimes \sum_{n=1}^{\infty} \frac{A(t)^{2n+1}}{\left( 2n+1 \right)!} \\
    & = I \otimes \cosh (A(t) ) + \sigma_z \otimes \sinh (A (t)) \, .
\end{align}
To determine the matrix elements of the reduced density matrix, we explicitly trace out the environmental degrees of freedom to focus upon
\begin{equation}
    \rho(t) = \operatorname{tr}_E \left\{ \tilde{U} (t) \rho (0) \otimes \rho_E \tilde{U}^{\dagger} (t) \right\}.
\end{equation}
From this, we can deduce that the coherences of the reduced density matrix evolve as
\begin{equation}
    \rho_{0 1}(t) = \rho_{ 0 1} (0)  \langle e^{2 A (t)}  \rangle \, ,
\end{equation} 
with $\rho_{ij}(t)=\bra{j}\rho(t)\ket{i}$ ($i,j=0,1$). Using the identity $\langle e^{A} \rangle = e^{\langle A \rangle^2/2}$, valid when the operator $A$ is a linear combination of creation and annihilation operators \cite{giuliani2005quantum}, we find that
\begin{align}
    \langle e^{ 2 A (t)}\rangle & = e^{- 2 \sum_k | \alpha_k (t) |^2 \langle b_k b_k^{\dagger} + b_k^{\dagger} b_k \rangle}  
     = e^{ - 2 \sum_k |\alpha_k (t)|^2 (2 N_k + 1 )} \, .
\end{align}
By substituting the expressions for $\alpha_k (t)$ and the mean occupation number $N_k$ of the $k^\text{th}$ mode of the environment, we obtain
$\langle e^{ 2 A (t)}\rangle = e^{- \Gamma (t)}$,
where we have assumed that the bath modes form a continuum. The function $\Gamma (t)$ is the decoherence function given in \Cref{eq:decoherencefunction} of the main text. Finally, transforming to the Schr{\"o}dinger picture, we recover \Cref{eq:Exactlysolvablerhot} of the main text. 

\section{Finding the Frequencies and Coupling Strengths of the RCs}
\label{app:C}

Here, we derive expressions for the frequencies $\Omega_i$, and coupling strengths $g_i$ of the RCs. By inserting the transformation in \Cref{eq:Bogoliubovtransformation} into the Hamiltonian before the mapping, as shown in \Cref{eq:Hamiltonian}, and then comparing the term that is linear in $X$ with the Hamiltonian after the mapping in \Cref{eq:Hamiltonianafter}, we find that the first $N$ columns of the orthogonal transformation have to be chosen such that 
\begin{equation}
    \sum_{p \leq N} \sqrt{\Omega_p} g_p \Lambda_{kp} = t_k \sqrt{\omega_k} \, .
\end{equation}
Comparing the terms with $b_i^{\dagger} b_i$ for $i \leq N$ yields
\begin{equation}
    \sum_{i \leq N, k} \Omega_i g_i^2 \Lambda_{ki}^2 = \sum_{i \leq N} g_i^2 \Omega_i \, .
\end{equation}
From these two expressions, we get
\begin{equation}
\sum_k t_k^2 \omega_k = \sum_{i \leq N} g_i^2 \Omega_i.
\end{equation}
Assuming a continuum of reservoir modes, we further find
\begin{equation}\label{eq:frequencycoupling1}
\frac{1}{g_i^2 \Omega_i^2} \int_0^{\infty} \textrm{d} \omega \, \omega J^L_i (\omega)=1.
\end{equation}
Comparing the terms that are quadratic in $X$ in the Hamiltonians gives
\begin{equation}
    \sum_{i \leq N} \frac{g_i^2}{\Omega_i} = \sum_k \frac{t_k^2}{\omega_k} \, ,
\end{equation}
Taking the continuum limit once more, we find
\begin{equation}\label{eq:frequencycoupling2}
    \frac{g_i^2}{\Omega_i} = \int_0^{\infty} \textrm{d} \omega \, \frac{J^L_i (\omega)}{\omega}  \, .
\end{equation}
Combining \Cref{eq:frequencycoupling1,eq:frequencycoupling2} we can deduce the frequency and coupling strengths of the RCs as shown in \Cref{eq:couplingstrengthRC}.

\section{Mapping of the Spectral Density}
\label{app:D}

For the mapping to be valid, the dynamics of the system resulting from the original and transformed Hamiltonians must be identical. Fortunately, it is not necessary to determine the orthogonal matrix $\Lambda$ directly, as it is fixed by knowledge of $J_{0}(\omega) = \sum_i J^L_i (\omega)$ alone. Our task is to find to establish a relationship between the $J^{RC}_i (\omega)$, which determine the dynamics of the system after the mapping, and the $J^L_i (\omega)$, which govern the dynamics before the mapping. To achieve this, we employ the Heisenberg equations of motion, which for an arbitrary system observable $O$ read as $\dot{O}^H = e^{iHt} \left[ H, O \right]e^{-i Ht}$, where $O^H = e^{iHt} O e^{-iHt}$. For the original representation, we obtain 
\begin{equation}
\begin{aligned}
   \dot{O}^H &= i S_1 + i S_2 \sum_k t_k \left( a_k^H + a_k^{\dagger H} \right),\\
    \dot{a}_k^H &= -i X^H t_k - i \omega_k a_k^H \, ,
    \end{aligned}
\end{equation}
with
\begin{equation}
    S_1 = e^{i H t} \left[ H_S{+}\sum_k \frac{t_k^2}{\omega_k} X^2 , O \right] e^{-i H t} \, ,
    S_2 = e^{i H t} [X, O] e^{-i H t} \, .
\end{equation}
We now apply the Fourier transform, $\hat{h}(z) = \int_{- \infty}^{\infty} \textrm{d}t \, h(t) e^{-i z t}$, to these equations, to obtain a set of coupled algebraic equations
\begin{equation}\label{eq:AOH}
\begin{aligned}
      z \hat{O}^H (z) &=  \hat{S}_1 (z) + \frac{1}{2 \pi} \int_{- \infty }^{\infty} \textrm{d} z'  \, \hat{S}_2 (z) \hat{S}_{3} (z-z'),\\
     z \hat{a}_k^H (z) &= - \omega_k \hat{a}_k^H (z) - t_k \hat{X}^H (z),
     \end{aligned}
\end{equation}
where
\begin{equation}\label{eq:S3z}
    \hat{S}_{3} (z) =  \sum_k t_k 
 \left(\hat{a}_k^H (z) + \hat{a}_k^{\dagger H} (z)\right) \, .
\end{equation}
We solve the last  of \Cref{eq:AOH} for $\hat{a}_k^H (z)$, deduce $\hat{a}_k^{\dagger H} (z)$ from it, and insert them into $\hat{S}_3(z)$, yielding
\begin{align}\label{eq:S4z}
    \hat{S}_3 (z) & = \sum_k \frac{2 \omega_k}{z^2 - \omega_k^2} t_k^2X^H (z)  = - \pi W_0 (z) X^H (z) \, ,
\end{align}
where we have taken the continuum limit and introduced the Cauchy transform of $J_0 (\omega)$ following the expressions
\begin{equation}
    W_n (z) = \frac{2}{\pi} \int_0^{\infty} \textrm{d} \omega \frac{\omega}{\omega^2 - z^2} J_n (\omega)= \frac{1}{\pi} \int_{- \infty}^{\infty} \textrm{d} \omega \frac{J_n (\omega)}{\omega - z}.
\end{equation}
The last equality sign holds provided that $J_n (\omega)$ is an odd function. Considering the Dirac delta function as the limit of a Lorentzian  \cite{alma991000282039708046}, changing variable as $z = \omega + i \epsilon$, and letting $\epsilon$ tend to $0$ from above, we have 
\begin{equation}\label{eq:cauchyprop}
    W_n (z) = \frac{1}{\pi} \mathcal{P} \int_{- \infty}^{\infty} \textrm{d} \omega' \frac{J_n (\omega')}{\omega' - \omega} + i J_n (\omega) \, .
\end{equation}

Similarly, we can derive the Heisenberg equations of motion in the mapped representation and use the Fourier transform to get  
\begin{equation}\label{eq:ATH}
     z \hat{O}^H (z) = \hat{S}_1 (z) + \sum_{i \leq N} \frac{g_i}{2 \pi } \int_{- \infty}^{\infty} \textrm{d} z' \, \hat{S_2} (z') \hat{S}_4 (z-z')  \, ,
\end{equation}
\begin{equation}\label{eq:bTH}
\begin{aligned}
     z \hat{b}_i^H (z) &= - \Omega_i \hat{b}_i^H (z) - g_i \hat{X}^H (z) {-}  2 \sum_{k > N } \frac{h_{ik}^2}{\Omega_k} \hat{S}_4 (z) {-} \hat{S}_{5} (z),\\
     z \hat{b}_i^{\dagger H} (z) &= \Omega_i \hat{b}_i^{\dagger H} (z) +  g_i \hat{X}^H (z) {+}  2 \sum_{k>N} \frac{h_{ik}^2}{\Omega_k} \hat{S}_4 (z) {+} \hat{S}_{5} (z),
     \end{aligned}
\end{equation}
\begin{equation}\label{eq:bkTH}
\begin{aligned}
    z \hat{b}_k^H (z) &= - \sum_{i \leq N} h_{ik} \hat{S}_4 (z) - \Omega_k \hat{b}_k^H (z),\\
    z \hat{b}_k^{\dagger H} (z) &= \sum_{i \leq N} h_{ik} \hat{S}_4 (z) + \Omega_k \hat{b}_k^{\dagger H} (z)
    \end{aligned}
\end{equation}
with
    $\hat{S}_4 (z) = \hat{b}_i^H (z) + \hat{b}_i ^{\dagger H} (z)$ and 
    $\hat{S}_{5} (z) = \sum_{k > N} h_{ik} \left( \hat{b}^H_k (z) + \hat{b}_k^{\dagger H} (z) \right)$.
We solve Equations~\eqref{eq:bkTH} for $\hat{b}_k^H (z)$ and $\hat{b}_k^{\dagger H} (z)$, subtract the first of Equations \eqref{eq:bTH} from the second, and then substitute the expressions for $\hat{b}_k^H (z)$ and $\hat{b}_k^{\dagger H} (z)$ into such difference to obtain
\begin{equation}\label{eq:S4Z}
    \hat{S}_4 (z) = \frac{2 g_i X^H (z)}{{z^2}/{\Omega_i} - \Omega_i+ 2 \pi W^L_i (z) - 4 \int_0^{\infty} \textrm{d} \omega {J^L_i (\omega)}/{\omega} }.
\end{equation}
Comparing \Cref{eq:ATH} with the corresponding expression for the original Hamiltonian in \Cref{eq:AOH}, using \Cref{eq:S4Z,eq:S3z} along with the property in \Cref{eq:cauchyprop}, we can infer the relation between the $J^{RC}_i (\omega)$'s and the $i^\text{th}$ contribution to $J_0 (\omega)$, as presented in \Cref{eq:SDmapping} of the main text.

\section{The Reaction Coordinate Master Equation}
\label{app:E}
We use \Cref{eq:BM} to derive a master equation that treats the coupling between the system and the RCs exactly, while the coupling with the residual environment is treated to second order within the Born-Markov approximation. In the Schr{\"o}dinger picture, the Born-Markov equation for the enlarged system is written as
\begin{table*}[t]
\centering
\begin{tabular}{c|ccccccc}
\hline
\multirow{2}{*}{\textbf{$\bm{\lambda_i}$ range}} & \multicolumn{7}{c}{\textbf{$\bm{\nu_i}$ range} (units of $\omega_0$)} \\ \cline{2-8} 
{(units of $\omega_0$)} & $ [ 0.54 , 0.57 )$ & $[0.57 , 0.7)$ & $[0.7, 0.75)$ & $[0.75, 0.88)$ & $[0.88,1.2)$ & $[1.2,1.8)$ & $\nu_i>1.8$ \\ \hline\hline
$(0.21,0.25]$ & $10$ & $8$ & $7$ & $6$ & $5$ & $4$ & $3$\\
 \hline
 $(0.2, 0.21]$ & $9$ & $8$ & $7$ & $6$ & $5$ & $4$ & $3$\\
\hline
 $(0.19, 0.2]$ & $9$ & $7$ & $7$ & $6$ & $5$ & $4$ & $3$\\
 \hline
 $(0.18, 0.19]$ & $9$ & $7$ & $6$ & $6$ & $5$ & $4$ & $3$\\
 \hline
 $(0.14, 0.18]$ & $8$ & $7$ & $6$ & $6$ & $5$ & $4$ & $3$\\
 \hline
 $(0.13, 0.14]$ & $7$ & $7$ & $6$ & $6$ & $5$ & $4$ & $3$\\
\hline
$(0.12, 0.13]$ & $7$ & $7$ & $6$ & $5$ & $5$ & $4$ & $3$\\
\hline
$(0.11, 0.12]$ & $7$ & $6$ & $6$ & $5$ & $5$ & $4$ & $3$\\
\hline
$(0.1, 0.11]$ & $7$ & $6$ & $5$ & $5$ & $5$ & $4$ & $3$\\
\hline
$\lambda_i = 0.1$ & $7$ & $6$ & $5$ & $5$ & $4$ & $4$ & $3$\\ \hline
\end{tabular}
\caption{Minimum required dimensions of the reaction coordinates (RCs) for simulating the pure dephasing model, based on the values of the Lorentzian peak parameters $\lambda_i$ and $\nu_i$. The dimensions were determined to ensure convergence to the exact solution and are suitable for any $\gamma_i$ within the range $[0.15, 0.25]\omega_0$. These dimensions are subsequently applied to the amplitude damping model, which is the main focus of our study.}
\label{tab:dimensions}
\end{table*}
\begin{align}
    \frac{\partial \rho_{S'} (t)}{\partial t}  & = - i \left[ H_{S'}, \rho_{S'} (t) \right] \nonumber \\
    & - \sum_{i \leq N} \int_{0}^{\infty} \textrm{d} \tau \left( C_i (\tau) \left[ A_i , e^{-i H_{S'} \tau} A_i e^{i H_{S'} \tau} \rho_{S'} (\tau) \right]  \right. \nonumber \\
    & \left. + C_i (- \tau) \left[ \rho_{S'} (t) e^{-i H_{S'} \tau} A_i e^{i H_{S'} \tau}, A_i \right] \right) \, ,
\end{align}
where $A_i = b_i + b_i^{\dagger}$, $\rho_{S'} (t)$ represents the state of the enlarged system, and $C_i (\tau)$ is the correlation function characterising the interaction between the $i$-th RC and the residual environment, which is defined in terms of $J^{RC}_i (\omega)$. Assuming that the residual environment is in a thermal Gibbs state, by changing the variable in the second term in \Cref{eq:bosoniccorrfunc} to $- \omega$, assuming that $J^{RC}_i (\omega)$ is an odd function, and using the fact that $n_B (-\omega) = -1 - n_B (\omega)$, we can express $C_i (\tau)$ as 
\begin{equation}
    C_i (\tau) = \int_{- \infty}^{\infty} \textrm{d} \omega J^{RC}_i (\omega) n_B (\omega) e^{i \omega \tau} \, .
\end{equation}
To calculate the one-sided Fourier transforms, we first find the eigenbasis for the Hamiltonian of the enlarged system, such that $H_{S'} \ket{\phi_j} = \psi_j \ket{\phi_j}$. This process allows us to write 
\begin{align}
    \int_0^{\infty} \textrm{d} \tau C_i (\tau) e^{-i H_{S'} \tau} A_i e^{i H_{S'} \tau} = \sum_{j,k} A^i_{jk} \Gamma^+_i (\nu_{jk}) \, ,
\end{align}
with $\nu_{jk} = \psi_j - \psi_k$, $A^i_{jk} = \bra{\phi_j} A_i \ket{\phi_k} \ket{\phi_j} \bra{\phi_k}$ and $\Gamma^+_i (\nu_{jk}) = \int_0^{\infty} \textrm{d} \tau C_i (\tau) e^{-i \nu_{jk} \tau}$. Explicitly calculating the integral with respect to $\tau$ yields
\begin{align}
 \Gamma^+_i (\nu_{jk}) & = \pi J^{RC}_i (\nu_{jk}) n_B (\nu_{jk}) - i \mathcal{P} \int_{- \infty}^{\infty} \textrm{d} \omega \frac{J^{RC}_i (\omega) n_B (\omega)}{\nu_{jk} - \omega}  \nonumber \\
 & \approx \pi J^{RC}_i (\nu_{jk}) n_B (\nu_{jk}) \, ,
\end{align}
where we have used the definition of the Dirac delta function as the limit of a Lorentzian function, and assumed that the imaginary part is negligible \cite{Strasberg_2016, 10.1063/1.4940218}. Care needs to be taken when evaluating at $\nu_{jk} = 0$, since $n_B (0)$ is divergent. To maintain generality and avoid committing to a specific form of the SD, we define $\Gamma^+_i (\omega)$ as presented in \Cref{eq:Gamma+} of the main text. Similarly, we define
\begin{equation}
    \int_0^{\infty} \textrm{d} \tau C_i ( - \tau) e^{-i H_{S'} \tau} A_i e^{i H_{s'} \tau} = \sum_{j,k} A^i_{jk} \Gamma^-_{jk} (\nu_{jk}) \, ,
\end{equation}
with $\Gamma^-_i (\omega)$ as shown in \Cref{eq:Gamma-} of the main text. With these definitions, we can rewrite the master equation in the form of \Cref{eq:RCME} in the main text.

\section{Dimensions of the Reaction Coordinates}
\label{app:F}

We analyse trajectories associated with SDs featuring one, two, or three Lorentzian peaks. For each Lorentzian component, we specify the parameter ranges as follows: $\gamma_i \in [0.15, 0.25]\omega_0$, $\lambda_i \in [0.1, 0.25]\omega_0$, and $\nu_i \in [0.54, 2]\omega_0$. The environmental temperature is set to $T = \omega_0/2$. To determine the optimal dimensions for the RCs across various parameter configurations, we first use the pure dephasing model as a benchmark. By simulating the dynamics of this model using the RC mapping, and comparing the results with the exact solution, we identify the minimum RC dimensions that ensure convergence within the RC mapping. 

The dimensions obtained through this benchmarking are shown in~\Cref{tab:dimensions}. We find that the optimal dimensions depend on $\lambda_i$ and $\nu_i$, but remain suitable for any $\gamma_i$ within the specified range. Once identified, these RC dimensions are applied to the amplitude damping model, which is the primary focus of our study. This approach is based on the assumption that the dimensions sufficient for accurately capturing the dynamics of the pure dephasing dynamics will also be adequate in the amplitude damping scenario.

\bibliography{biblio.bib}

\end{document}